# First principles derivation of a Rayleigh-Gans-Debye model for scattering from anisotropic inhomogeneities

M. H. Shachar and J. E. Garay*

Materials Science & Engineering Program, Mechanical & Aerospace Engineering Department
University of California San Diego
La Jolla, California
*Correspondence: jegaray@ucsd.edu

**Abstract**

Scattering problems are important in describing light propagation in wide ranging media such as the atmosphere, colloidal solutions, metamaterials, glass ceramic composites, transparent polycrystalline ceramics, and surfaces. The Rayleigh-Gans-Debye (RGD) approximation has enjoyed great success in describing a wide range of scattering phenomena. We derive a generalized RGD formulation from the perturbation of Maxwell's equations. In contrast to most treatments of RGD scattering, our formulation can model any soft scattering phenomena in linear media, including scattering by stochastic process, lossy media, and by anisotropic inhomogeneities occurring at multiple length scales. Our first-principles derivation makes explicit underlying assumptions and provides jumping-off points for more general treatments. The derivation also facilitates a deeper understanding of soft scattering. It is demonstrated that sources of scattering are not interfaces as is often presumed, but excess accelerating charges emitting uncompensated radiation. Approximations to the equations are also presented and discussed. For example, the scattering coefficient in the large-size RGD limit is shown to be proportional to the correlation length and the variance of a projected phase shift.

**Contents**

**TABLE 1**: List of Symbols

| Symbol | Name |
|---|---|
| $\boldsymbol{E}$ | Electric field; Electric field phasor |
| $\nabla \times$ | Curl operator |
| $\mu$ | Relative permeability |
| $\epsilon$ | Relative susceptibility tensor |
| $k_0$ | Wavenumber of light in vacuum |
| $\mu_0$ | Vacuum permeability |
| $\epsilon_0$ | Vacuum susceptibility |
| $\omega$ | Angular frequency |
| $\lambda_0$ | Vacuum wavelength |
| $\boldsymbol{r}$ | Spatial coordinate |
| $\bar{\epsilon}$ | Averaged permittivity tensor |
| $\bar{\boldsymbol{E}}$ | Incident field |
| $\bar{E}_0$ | Unidirectional incident field amplitude |
| $\widehat{\boldsymbol{E}}_0$ | Unidirectional incident polarization direction |
| $\boldsymbol{k}$ | Wavevector of incident light |
| $\widetilde{\boldsymbol{E}}$ | Electric scatter field due to inhomogeneities |
| $\tilde{\epsilon}$ | Permittivity fluctuation tensor |
| $\boldsymbol{Q}$ | Scattering source term |
| $\sigma$ | Conductivity tensor |
| $\boldsymbol{J}$ | Electric current density; Electric current density phasor |
| $t$ | time |
| $\langle \cdot \rangle_W$ | Sample ensemble average |
| $\langle \cdot \rangle_V$ | Volume average |
| $\|\cdot\|$ | Linear operator norm |
| $\lambda_{max}(A)$ | Max eigenvalue of $A$ |
| $\dagger$-superscript | Denotes adjoint; conjugate transpose |
| $\mathrm{tr}(A)$ | Trace of $A$ |
| $\Delta\phi$ | Phase difference between unperturbed and perturbed solutions |
| $L_c$ | Correlation length of scattering sources |
| $\tilde{n}_{max}$ | Largest index difference |

| Symbol | Description |
|---|---|
| $m$ | Relative refractive index |
| $x$ | Size parameter of scatterer |
| $\Gamma(\boldsymbol{r}, \boldsymbol{r}')$ | Dyadic Green's function at $\boldsymbol{r}$ for source at $\boldsymbol{r}'$ |
| $\delta(\boldsymbol{r})$ | 3D Dirac-delta function |
| $\mathcal{V}$ | 3D vector space; a normed vector space |
| $I_d$ | Identity tensor; Identity matrix |
| $\otimes$ | Complex dyadic product |
| $\bar{n}$ | Root mean square of refractive index |
| $\langle \cdot \rangle_\Omega$ | Average along all crystallographic orientations |
| $n_{ff}$ | Far-field refractive index |
| $\Re$ | Real part of complex number |
| $I$ | Intensity (irradiance) |
| $\beta$ | Angle between Poynting vector and illuminated surface normal |
| $\bar{n}_{ff}$ | Orientation averaged far-field refractive index |
| $c$ | Speed of light in vacuum |
| $G$ | Intensity proportionality constant |
| $\tilde{I}$ | Scattering intensity |
| $\boldsymbol{r}', \boldsymbol{r}''$ | Locations of scattering sources |
| $\mathcal{v}$ | Set of points in the volume of the scatterer |
| $1_\mathcal{v}(\boldsymbol{r})$ | Scattering volume indicator function |
| $\boldsymbol{q}$ | Momentum transfer |
| $\boldsymbol{\rho}$ | Vector difference between two impulse point sources |
| $\hat{r}_\perp$ | Vector rejection from $\hat{\boldsymbol{r}}$ operator |
| $I_0$ | Intensity of incident light |
| $K$ | Scattering source covariance function for unit electric field wave |
| $\mathcal{F}$ | Fourier transform |
| $\mathcal{F}_x$ | Fourier transform on variable $x$ |
| $\mathcal{F}^{-1}$ | Inverse Fourier transform |
| $\star$ | Convolution operator |
| $V$ | Volume of the scattering medium |
| $\mathcal{v}_I$ | Set of points in the interior volume of the scatterer |
| $\mathcal{v}_S$ | Set of points in the surface volume of the scatterer |
| $|\mathcal{v}_I|$ | Volume of interior points |
| $|\mathcal{v}_S|$ | Volume of surface points |
| $\hat{\boldsymbol{\rho}}$ | Unit vector parallel to $\boldsymbol{\rho}$ |
| $\boldsymbol{Q}_\perp$ | Vector rejection of source term from the observation direction |
| $k$ | Wavenumber of incident light |
| $Q_\perp$ | Magnitude of $\boldsymbol{Q}_\perp$ |
| $\text{Var}_W$ | Sample ensemble variance |
| $K_0$ | Scattering source strength |
| $\kappa$ | Scattering source autocorrelation |
| $\kappa_{\bar{E}}$ | Incident light amplitude autocorrelation |
| $\kappa_W$ | Structure autocorrelation |
| $\tau_{\bar{E}}$ | Coherence time of incident light |

| Symbol | Description |
|---|---|
| $L_{\bar{E}}$ | Coherence length of incident light |
| $W$ | Set of samples produced by a random process |
| $\rho$ | Magnitude of $\boldsymbol{\rho}$ |
| $q(\hat{\boldsymbol{r}})$ | Magnitude of $\boldsymbol{q}$ in the direction $\hat{\boldsymbol{r}}$ |
| $\varphi$ | Angle between $\boldsymbol{k}$ and $\boldsymbol{r}$ |
| $\hat{\mathcal{V}}$ | Set of all unit vectors in vector space $\mathcal{V}$ |
| $\hat{\boldsymbol{r}}$ | Unit vector parallel to $\boldsymbol{r}$ |
| $\alpha_{sc}$ | Total scattering intensity |
| $A$ | Cross sectional area |
| $\ell$ | Optical path length |
| $\mu_{sc}$ | Scattering coefficient |
| $L_W$ | Coherence length of inhomogeneities |
| $K'$ | Proportionality constant |
| $n$ | Index of refraction (generally a tensor) |
| $\mathfrak{I}$ | Imaginary component of complex number |
| $*$-superscript | complex conjugate |
| $\langle \cdot \rangle_{\hat{r}}$ | Average along all observation directions |
| $V_{corr}$ | Correlation volume |
| $\hat{\boldsymbol{k}}$ | Unit vector parallel to $\boldsymbol{k}$ |
| $\mathcal{V} \perp \hat{\boldsymbol{k}}$ | All vectors perpendicular to $\boldsymbol{k}$ |
| $\delta_k(\boldsymbol{q})$ | Dirac surface delta for sphere of radius $k$ |
| $\delta(r)$ | One-dimensional Dirac delta function |
| $\langle \boldsymbol{v}, \boldsymbol{w} \rangle$ | Inner product of vectors $\boldsymbol{v}, \boldsymbol{w}$ |
| $L_{corr}(\hat{\boldsymbol{k}})$ | Correlation length in the direction $\hat{\boldsymbol{k}}$ |
| $\hat{k}_{\perp}$ | Vector rejection from $\hat{\boldsymbol{k}}$ operator |
| $\dfrac{d\phi_{\perp}}{dl}$ | Delay coefficient of light traveling in the $\hat{k}$ direction |

## 1. Introduction

Light scattering due to inhomogeneities is ubiquitous in nature and technology. Important examples are light scattering by particles in the atmosphere [1] and in colloidal materials [2]. Assessing light scattering is crucial in developing novel optical materials such as optical metamaterials [3-6], glass-ceramic composites [7] and transparent polycrystalline ceramics [8,9]. Our overall goal in this series of two papers is to develop analytical approaches for treating scattering in inhomogeneous dielectrics. We believe the expressions derived will be useful for experimentalists designing and testing new optical materials as well as for scientists interested in further model development. Benefits include clearly laid out assumptions leading to analytical equations that can serve as jumping off points for future analytical or numerical work.

We develop an approach for treating elastic electromagnetic scattering due to weak inhomogeneities. We believe the approach taken here with clearly laid out assumptions will be useful for connecting optical properties with material design. In particular, expressions derived here should be useful for treating contemporary material development such as biological materials [2,3], glass-ceramic composites [4,5] and polycrystalline optical ceramics [6-8].

While the resulting equations are important, most have been arrived at in some form previously in the literature [9-24]. In 1881, Rayleigh [9] used a first-order perturbation of Maxwell's equations to describe the scattering of light by weak, scalar inhomogeneities of permittivity and permeability. This result was later rederived by Gans [10]. Rayleigh, In the same paper, applied his results to the analysis of a sphere (and to an infinite cylinder) of uniform disturbance. In 1910, Einstein [11] used a first-order perturbation of Maxwell's equations to describe scattering by a fluid that is near the critical point. The permittivity fluctuations were expressed in terms of density fluctuations and those were described using statistical mechanics. In 1914, Brillouin [12] applied Einstein's results to density fluctuations caused by thermal fluctuations. In 1949, Debye and Bueche [13] used Einstein's results to develop a technique to characterize inhomogeneities using scattering measurements. Debye and Bueche also re-express the intensity equations using a correlation function. In all these works, the inhomogeneities were always described using a scalar permittivity. In 1955, Goldstein and Michalik [14] extended Debye and Bueche's results to non-absorbing inhomogeneities which have uniaxial symmetry. In 1980, Ross and Nieto-Vesperinas [15] published a paper generalizing the problem to absorbing media. In the same year, Ross and Jarvis published a series of two papers [16, 17] that generalize the problem to anisotropic media.

Many of the results discussed above will be reproduced here. With that in mind, the goal of this paper is four-fold:

(1) provide a clear first-principles derivation

(2) maintain the generality of the equations for as long as possible

(3) impart the physical interpretation of important results

(4) reformulate results to make them more readily applicable for comparison with experiment

The first goal is accomplished by making explicit all assumptions, definitions, and techniques – particularly those that are often taken for granted such as the bulk approximation and the stochastic elements of the theory. The second goal is accomplished by refraining from employing assumptions until necessary. As a result, most of the analysis remains general enough to accommodate an anisotropic permittivity field and absorption. The explicit mention of the assumptions creates natural jumping-off points for future researchers interested in applying these techniques to problems which violate of some of the assumptions. The third goal is accomplished by re-expressing equations such that their interpretation becomes clearer and then providing that interpretation. The fourth goal is accomplished by expressing scattering using a scattering coefficient, which can be extracted from in-line transmission measurements [25].

Although not the primary objective of the paper, some novel results will be showcased. In particular the additional step of calculating scattering coefficient is performed for many of the results, making them more readily available for in-line transmission models which are often used to compare with experimental results. In addition, the effect of the scattering volume geometry on the scattering intensity is provided in **section 3.2**. Furthermore, a generalized equation for the large-size Rayleigh-Gans-Debye (RGD) scattering coefficient is provided in **section 3.7.4**.

We apply a first-order perturbation approach to the scattering equation to solve for scattered field. We then find the far-field scattering intensity caused by the perturbation field. Since it is often impractical to provide the microscopic properties (e.g. permittivity and permeability) of a sample at every point, we treat the scattering source term stochastically. Using this method we provide expressions that describe the total scattering intensity caused by a small scattering volume.

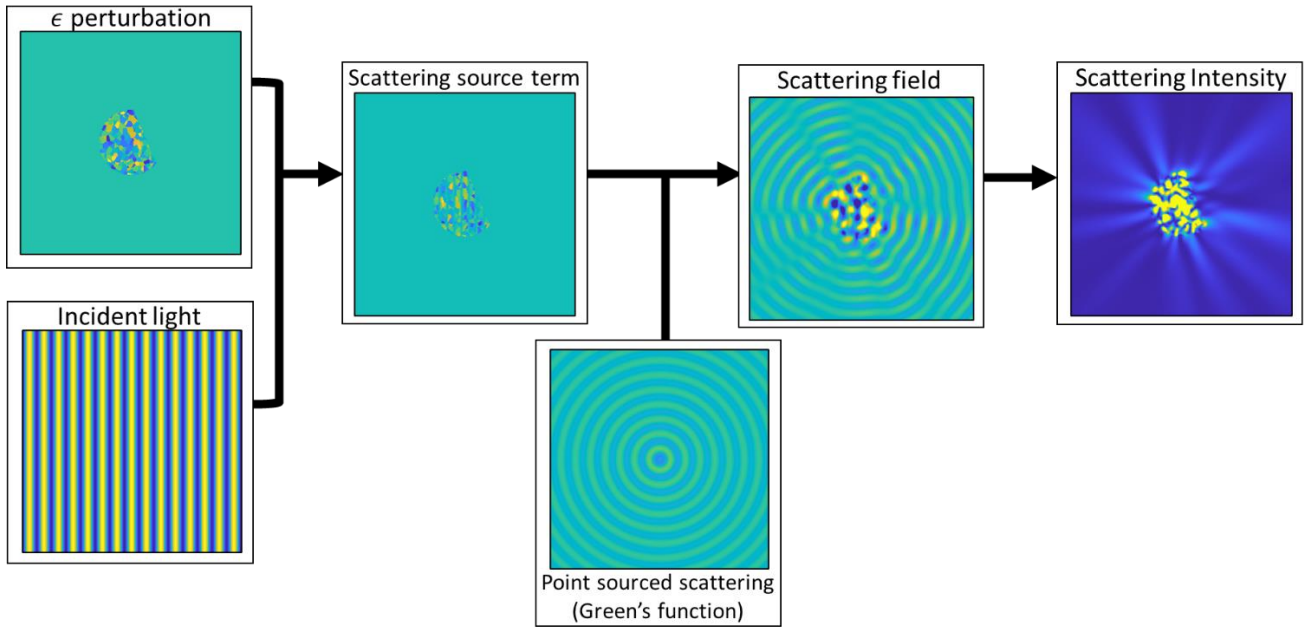

**Figure 1:** Schematic overview of a procedure for calculating inhomogeneous scattering intensity

**Figure 1** provides a schematic overview of the analysis used to solve for the soft scattering intensity due to inhomogeneity. Information about the inhomogeneity is provided in the form of a space dependent $\epsilon$ function. In addition, information about the incident light is provided via its wavelength and propagation direction (or alternatively, by its wavevector). Together, the inhomogeneity and incident light can be used to compute the scattering source term. The scattering source term is proportional to the excess accelerating charge induced by inhomogeneities in the scattering volume. Exploiting linearity, the source is broken up into impulses and their far-field responses (given by a far-field Green's function), which are then summed to produce the overall far-field scattering field. Finally, the field is squared to provide a quantity proportional to the total scattering intensity. To compute the scattering coefficient, the scattered field intensity in all orientations can be integrated.

The procedure in **Figure 2** extends the procedure in **Figure 1** to an *ensemble* of scattering volumes produced by a random process. Information about the inhomogeneity is now provided in the form of a random variable $\epsilon$. When the random process is space invariant, the random variable $\epsilon$ is independent of the location in space. Treating the material stochastically makes $\epsilon$ far easier to specify and model, but complicates the determination of the scattering intensity. **Figure 2** provides a schematic overview of a stochastic approach for determining scattering intensity which mirrors the process in **Figure 1**. The process in is now carried out on the ensemble, resulting in an ensemble of far-field scattering intensities. The ensemble is then averaged to produce an averaged far-field scattering intensity, which can then be integrated to produce the scattering coefficient for the random process.

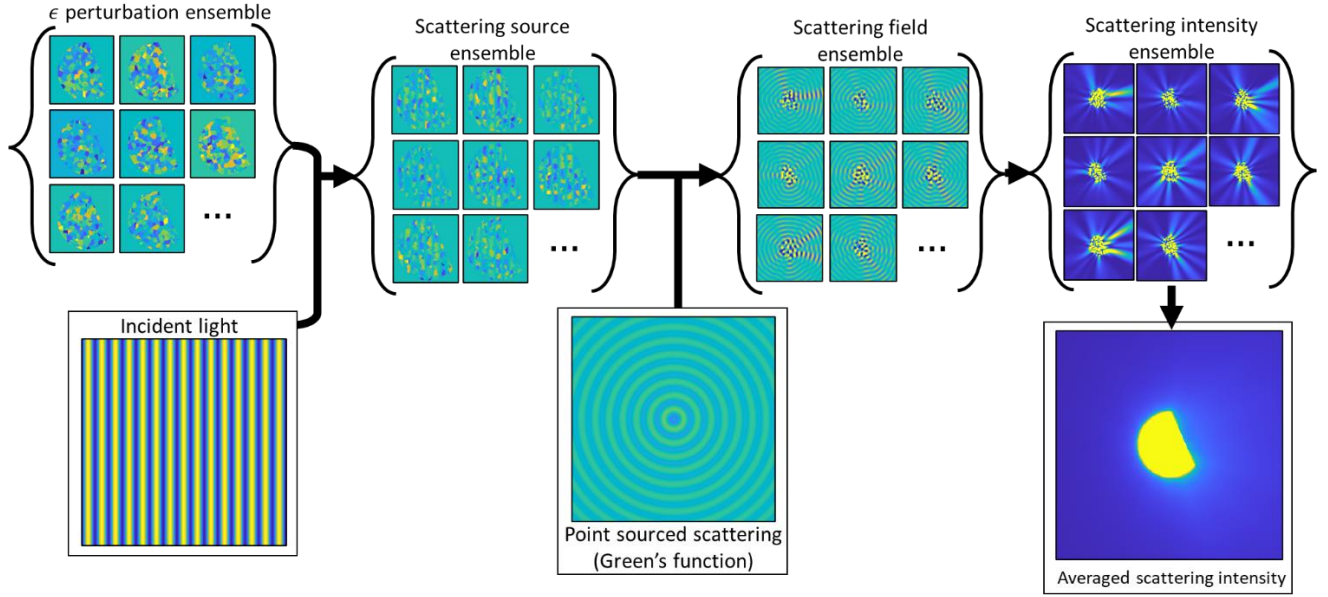

**Figure 2:** Schematic overview of a stochastic approach for calculating inhomogeneous scattering intensity

The methods outlined in **Figures 1 and 2** result in analytic expressions for the scattering coefficient. We then use our expressions to explore approximations for various size limits: The Rayleigh small-size limit approximation and the large-size Rayleigh-Gans-Debye (RGD) limit.

There are several advantages of this approach. Our first principles derivation from Maxwell's equations facilitates a deeper understanding of the source of inhomogeneous scattering. For example the derived expressions clearly demonstrate that the sources of scattering in inhomogeneous media are excess accelerating charges and not, as is commonly supposed, index contrast in the medium ( see, for instance, the frequent claim that grain boundaries scatter in polycrystals [26-36]). In addition, our approach produces multiple jumping-off points for the development of more general models. For instance, non-soft scattering and non-weak anisotropy can be approached by following the derivation up to the point where soft scattering and weak anisotropy are assumed. Then, for non-soft scattering a higher order perturbation might be employed while for non-weak anisotropy, the anisotropic dyadic Green's function may be used.

## 2. Soft scattering field in dielectrics

The goal is to describe elastic electromagnetic scattering in materials with weakly non-uniform optical properties (in a way that will be made precise in **section 2.3**). In an ohmic, time-invariant material the scattering fields for monochromatic incident light can be described by the equation

$$\nabla \times (\nabla \times \boldsymbol{E}) - \mu\epsilon k_0^2 \boldsymbol{E} = 0 \qquad (1)$$

Where $\nabla \times$ is the curl, $\mu$ is the relative permeability, $\epsilon$ is the relative permittivity, $k_0$ is the vacuum wavenumber, and $\boldsymbol{E}$ is the electric field. See **appendix A (Supplementary information)** for a derivation of this equation from Maxwell's equations. The vacuum wavenumber is related to the angular frequency of the incident light by

$$k_0^2 := \mu_0 \epsilon_0 \omega^2 \tag{2}$$

and to the vacuum wavelength $\lambda_0$ by

$$k_0 = 2\pi/\lambda_0 \tag{3}$$

Nearly all dielectrics have a weak magnetic response making the non-magnetic approximation

$$\mu \approx 1 \tag{4}$$

valid. If only non-magnetic media are considered, the equation that governs the electric field then becomes

$$\nabla \times (\nabla \times \mathbf{E}) - \epsilon k_0^2 \mathbf{E} = 0 \tag{5}$$

Where, in general, $\epsilon$ is a tensor-valued function of space. We will be solving equation (5) for inhomogeneities of $\epsilon$. However, the solution of the slightly more general equation (1), which includes inhomogeneities of $\mu$, can be recovered by replacing $\epsilon$ with $\mu\epsilon$ everywhere in the results.

## 2.1 First-order perturbation of the scattering equation

Equation (5) is a PDE with variable coefficients when the electrical permittivity $\epsilon$ is a function of position $\mathbf{r}$. There is no known general analytic technique for an exact solution to such equations, but a perturbation approach can give approximate solutions from a similar equation whose solutions are known. A solution for which an exact solution is known is

$$\nabla \times (\nabla \times \mathbf{E}) = k_0^2 \bar{\epsilon} \mathbf{E} \tag{6}$$

Where the non-uniform permittivity $\epsilon$ has been replaced by a uniform, average permittivity $\bar{\epsilon}$. The exact solution to (6) will be denoted $\bar{\mathbf{E}}$ and can be thought of as the incident field. The solution to (6) is a linear combination of the unidirectional solutions, each of which is of the form

$$\begin{aligned}\bar{\mathbf{E}} &= \bar{E}_0 \widehat{\mathbf{E}}_0 \exp(i\mathbf{k} \cdot \mathbf{r}) \\ |k^2| &= |\mathbf{k} \cdot \mathbf{k}| = k_0^2 |\bar{\epsilon} \widehat{E}_{0,k}|\end{aligned} \tag{7}$$

The complex amplitude is represented by $\bar{E}_0$, the polarization by $\widehat{\mathbf{E}}_0$, the wavenumber in the medium is $k$, and the wavevector by $\mathbf{k}$. Every choice of $\mathbf{k}$ is linearly independent from every other choice of $\mathbf{k}$. In addition, there are two linearly independent polarization axis to choose from. Any linear combination of these solutions will produce another solution. The appropriate choice of solution will depend on the boundary conditions imposed by the incident radiation. The scattered field due to inhomogeneities $\widetilde{\mathbf{E}}$ can be written as

$$\widetilde{\mathbf{E}} = \mathbf{E} - \bar{\mathbf{E}} \tag{8}$$

Since $\widetilde{\mathbf{E}}$ is approximated by difference between the perturbed field $\mathbf{E}$, which approximates the solution to (5), and the incident field $\bar{\mathbf{E}}$. It is also useful to define the fluctuation of permittivity around the mean

$$\tilde{\epsilon} := \epsilon - \bar{\epsilon} \tag{9}$$

Assuming a first order perturbation is a good enough approximation (the conditions under which this occurs will be analyzed below in **section 2.3**), the PDE describing the scattering field can be expressed as

$$\nabla \times (\nabla \times \widetilde{\mathbf{E}}) - k_0^2 \bar{\epsilon} \widetilde{\mathbf{E}} = \mathbf{Q} \tag{10}$$

Which is an inhomogeneous PDE where the source term is

$$\mathbf{Q}(\mathbf{r}) = k_0^2 \tilde{\epsilon} \bar{\mathbf{E}} \tag{11}$$

The source term represents the sources of scattering. Solving equation (10) for $\widetilde{\boldsymbol{E}}$ will provide a first order approximation of the scattering due to inhomogeneities. This source term can be shown to be proportional, approximately, to the density of excess accelerating charge at $\boldsymbol{r}$. The complex electrical conductivity, permittivity, and permeability are related by

$$\mu\epsilon := 1 + \frac{i\sigma}{\omega\epsilon_0} \tag{12}$$

Using the non-magnetic approximation $\mu \approx 1$

$$\sigma = i\omega\epsilon_0(1-\epsilon) \tag{13}$$

Since the current density is proportional to the velocity of charges the accelerating charge is proportional the time derivative of the current density which can be expressed as

$$\frac{\partial \boldsymbol{J}}{\partial t} = -i\omega \boldsymbol{J} = -i\omega\sigma \boldsymbol{E} = \omega^2 \epsilon_0(1-\epsilon)\boldsymbol{E} = \frac{k_0^2}{\mu_0}(1-\epsilon)\boldsymbol{E} \tag{14}$$

As discussed in **section 2.3**, the validity of the first order perturbation relies on the electric field being close to the incident field. Assuming $\boldsymbol{E} \approx \overline{\boldsymbol{E}}$, the excess accelerating charge can be expressed as

$$\frac{\partial \boldsymbol{J}}{\partial t} - \frac{\partial \overline{\boldsymbol{J}}}{\partial t} = -\frac{k_0^2}{\mu_0}\tilde{\epsilon}\overline{\boldsymbol{E}} \tag{15}$$

Comparing this result to equation (11) demonstrates that the scattering source term is proportional to the excess accelerating charges.

**2.2 Stochastic source term**

The far-field scattering intensity is the intensity from the perturbation field $\widetilde{\boldsymbol{E}}$. Equation (10) is an equation whose solution will approximate $\widetilde{\boldsymbol{E}}$ (see **section 2.3** for a criterion for the validity of the approximation). However, to formulate equation (10), the source term $\boldsymbol{Q}$ must be fully specified. To do so the source term, and thus the permittivity $\epsilon(\boldsymbol{r})$ at every point $\boldsymbol{r}$ inside the scattering sample is necessary. Knowledge of this kind is impractical to furnish. For a birefringent polycrystal, for example, this would require knowing the crystallographic alignment of every grain in addition to the 3D morphology of each grain. In addition, it is not usually of practical interest to know the exact scattering profile that a specific sample will produce. Instead, it is much more useful to describe the scattering from inhomogeneities produced from a known random process, such as by a known materials synthesis/processing process. Examples of commonly employed synthesis processed for dielectrics are melt casting [37], fiber drawing [37], powder sintering/densification [6-8] and self-assembly [38]. In addition samples manufactured using a particular technique can have their microscopic state altered by applied fields for example stress and temperature gradients.

For generality we will refer to these simply as a random process. Such a process will produce an ensemble $W$ of samples, each varying from the other by their microscopic state but having similar macroscopic properties. Our goal, then, will be to describe the scattering of this ensemble.

To make any progress, the assumption of ensemble ergodicity is required. Ensemble ergodicity states that *as more samples from a process are measured, the average of their scattering profiles will tend toward the ensemble average.* A necessary condition for sample ergodicity is shown in **Figure 3.** As the sample size of the ensemble is increased, the sample average of the optical properties at each point will approach the ensemble average of the optical properties. Violation of this assumption will lead to unpredictable scattering from samples produced by the random process. In addition to being intractable, this would make the process under investigation impractical as well.

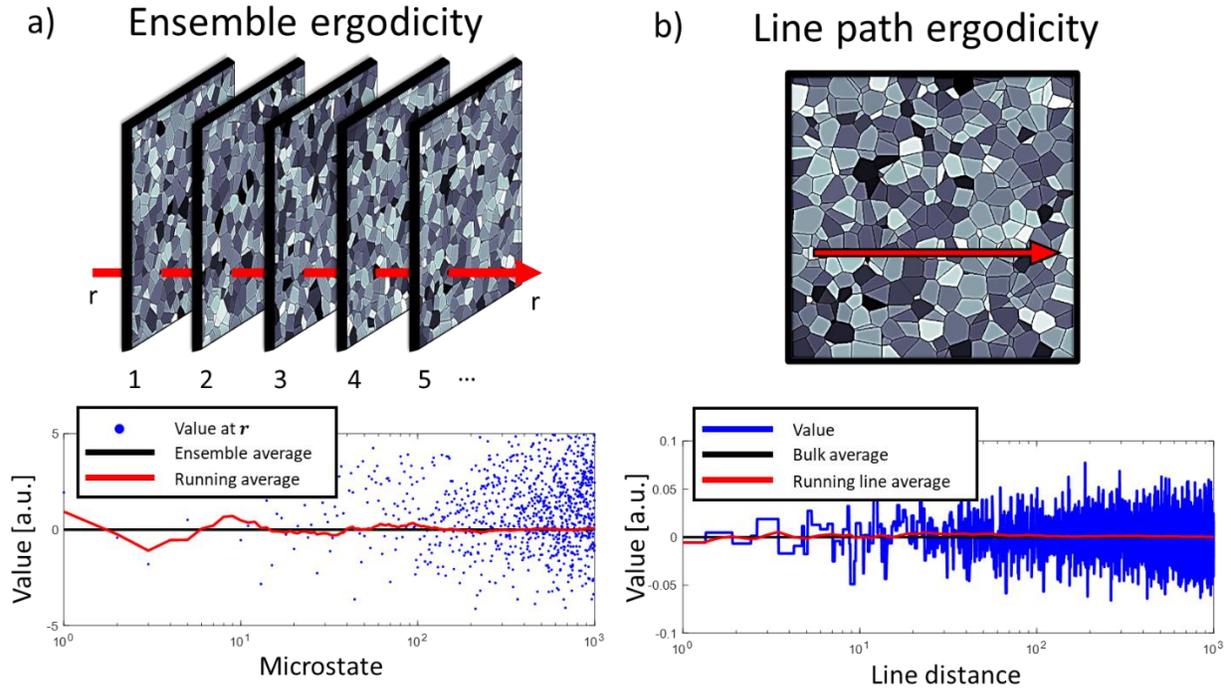

**Figure 3:** Schematics of ergodicity in a) Ensemble and b) Line path

With the assumption of ergodicity, it becomes possible to compute the average behavior of a random process. Mathematically, the permittivity fluctuation profile $\tilde{\epsilon}(r)$ will be a random variable representing the permittivity fluctuation of a sample produced by the random process. Using the stochastic framework, it is useful to redefine the average permittivity first introduced in equation (6) as

$$\bar{\epsilon} \coloneqq \langle\langle\epsilon\rangle_W\rangle_V \tag{16}$$

Where $\langle\cdot\rangle_W$ is the ensemble average and $\langle\cdot\rangle_V$ is the volume average. The advantage of using (16) is that $\bar{\epsilon}$ will not be a function of the microstate. Another assumption that we will find useful to employ is the spatial invariance of the random process, or simply space invariance (not to be confused with uniformity).

**Figure 4** schematically contrasts space invariance, uniformity/homogeneity, and non-uniformity/inhomogeneity. As a consequence of space invariance, the statistics of $\epsilon(r)$ cannot explicitly depend on the coordinate $r$. In such a case, the average permittivity simplifies to

$$\bar{\epsilon} = \langle\epsilon\rangle_W \tag{17}$$

If the evolution of $\epsilon(r)$ along straight lines is also ergodic (as presented in **Figure 3** under line path ergodicity), then for volumes large relative to the correlation length $L_C$ of fluctuations (the bulk limit as described in **section 3.3**)

$$\bar{\epsilon} \approx \langle\epsilon\rangle_V \tag{18}$$

Equation (18) gives a practical way to approximate $\bar{\epsilon}$ from a single, sufficiently large sample. Note that, in our usage of the terms, a spatially invariant random process can produce a non-uniform permittivity. In general, we will reserve the terms "uniform" and "homogeneous" for properties or functions that are constant in space and "space invariant" for random processes that treat all locations in space as equivalent (processes for which the statistics are unform).

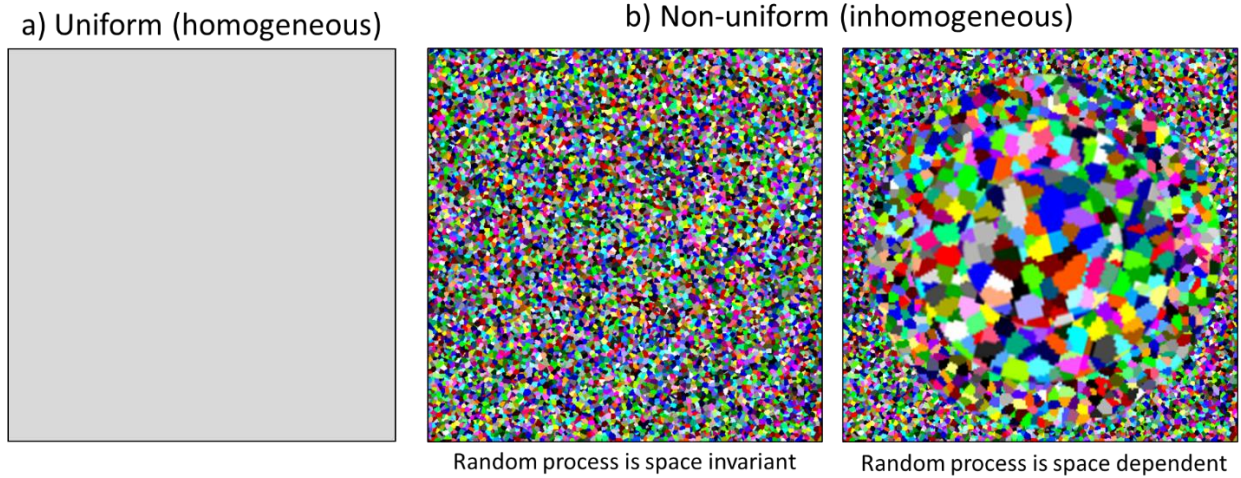

**Figure 4:** Schematic depictions of a) space invariant uniformity/homogeneity, and b) non-uniformity/inhomogeneity. In the inhomogeneous case the random process may be space dependent or space invariant.

### 2.3 Validity of first-order perturbation

For the first-order perturbation (10) to be a valid approximation, the deviation in magnitude and phase of the perturbed fields and the incident fields should be small. For this, two assumptions are sufficient: (1) soft scattering and (2) a small phase difference.

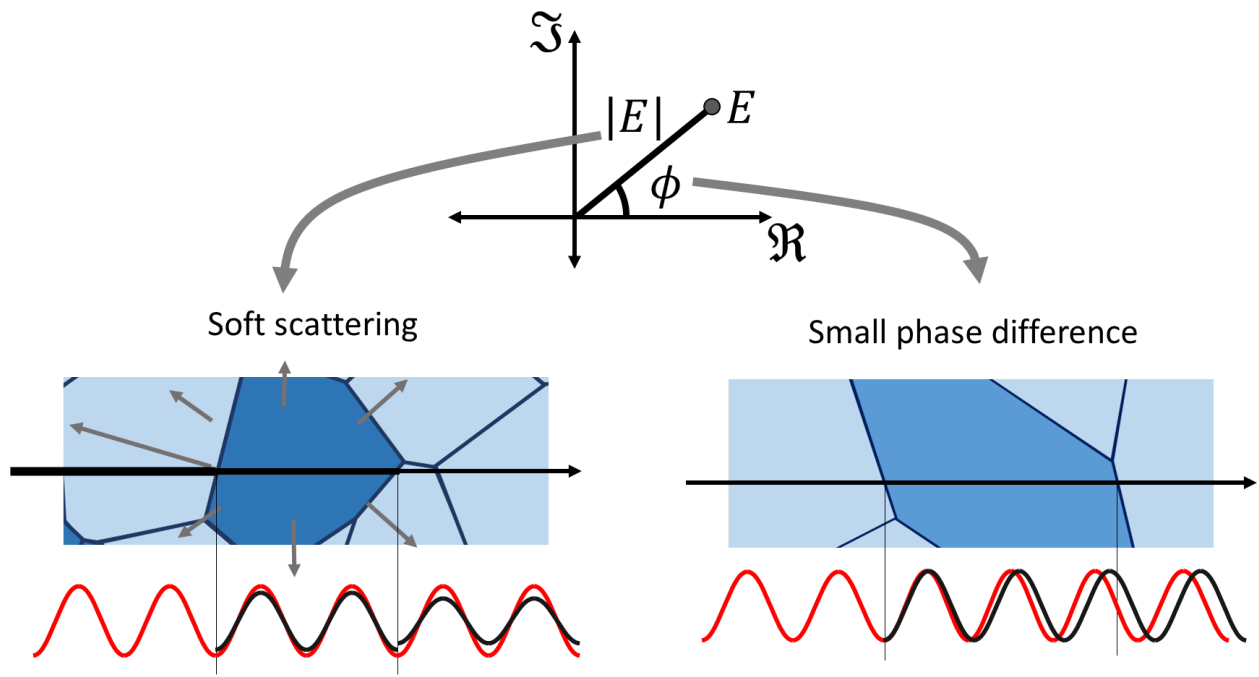

**Figure 5:** Schematic representation of the two assumptions (soft scattering and Small phase difference) necessary for the first order perturbation approximation to be valid

**Figure 5** schematically illustrates the two assumptions. Enforcing a small deviation in magnitude results in the soft scattering assumption. A large enough index variation will lead to significant losses over a correlation length. If these losses are large enough, the magnitude of the perturbed and incident fields will be different enough to require higher order perturbation terms. Similarly, enforcing a small deviation in phase leads to the small phase difference approximation. If the index variation or the correlation length are large enough, the phase difference accumulated over a correlation length will be too large. Once again, higher order perturbation terms will be required to account for these phase delays.

A mathematical description of the two assumptions will make them more precise. The soft scattering assumption is satisfied when the strength of the inhomogeneities is small. The strength of the inhomogeneities can be quantified using a generalization of the linear operator norm as defined in **appendix B**

$$\|\tilde{\epsilon}\| = \sqrt{\langle \lambda_{max}(\tilde{\epsilon}^\dagger \tilde{\epsilon})\rangle_W} \quad (19)$$

where $\langle \cdot \rangle_W$ is the ensemble average. If $\tilde{\epsilon}$ is Hermitian, which is true for passive, lossless media (see **appendix C**), then

$$\|\tilde{\epsilon}\| = \sqrt{\langle |\lambda_{max}(\tilde{\epsilon})|^2\rangle_W} \quad (20)$$

in which case the norm is the largest permittivity. The generalization of the norm is to allow for a distribution. If the operator is a scalar, then the operator simplifies further to

$$\|\tilde{\epsilon}\| = \sqrt{\langle |\tilde{\epsilon}|^2\rangle_W} \quad (21)$$

which is simply the RMS of $\tilde{\epsilon}$. Using the operator norm, we may express the soft scattering assumption more precisely as

$$\|\tilde{\epsilon}\| \ll \frac{1}{3}|\text{tr}(\bar{\epsilon})| \quad (22)$$

where $\text{tr}(\cdot)$ is the trace of an operator. If $\bar{\epsilon}$ is isotropic (a scalar), then (22) simplifies to

$$\|\tilde{\epsilon}\| \ll |\bar{\epsilon}| \quad (23)$$

The second assumption required for the first-order perturbation to be a valid approximation is the small phase difference assumption. It is satisfied when the phase difference $\Delta\phi$ caused within a scatterer is much smaller than a full cycle. A reasonable criterion for whether $\Delta\phi$ is small is

$$\Delta\phi \ll 1 \quad (24)$$

An upper estimate of the index difference that can occur is

$$\tilde{n}_{max} \coloneqq \sqrt{\|\tilde{\epsilon}\|} \quad (25)$$

If $L_c$ denotes the correlation length of the inhomogeneities, then the phase difference caused by fluctuations of length $L_c$ and index difference $\tilde{n}_{max}$ is

$$\Delta\phi = k_0 L_c \tilde{n}_{max} \quad (26)$$

And the small phase difference criterion becomes

$$k_0 L_c \tilde{n}_{max} \ll 1 \quad (27)$$

Equation (10) and (11) comprise the Rayleigh-Gans-Debye approximation (RGDA) seen throughout scattering and diffraction literature. Conditions (24) and (27) make up the assumptions typically required to employ RGDA. These conditions are usually expressed as

$$|m - 1| \ll 1 \tag{28}$$

$$x|m - 1| \ll 1 \tag{29}$$

Where $m$ is the relative refractive index between scatterer and medium and $x$ is the size parameter of the scatterer. Condition (24) corresponds to condition (28) and allows us to ignore the effect that inhomogeneities have on the propagation of scattered light. Condition (27) corresponds to condition (29) and ensures that the scattering is soft enough that the incident radiation is a good enough approximation for use in the source term as described in equation (11).

**2.4 Solution using dyadic Green's function**

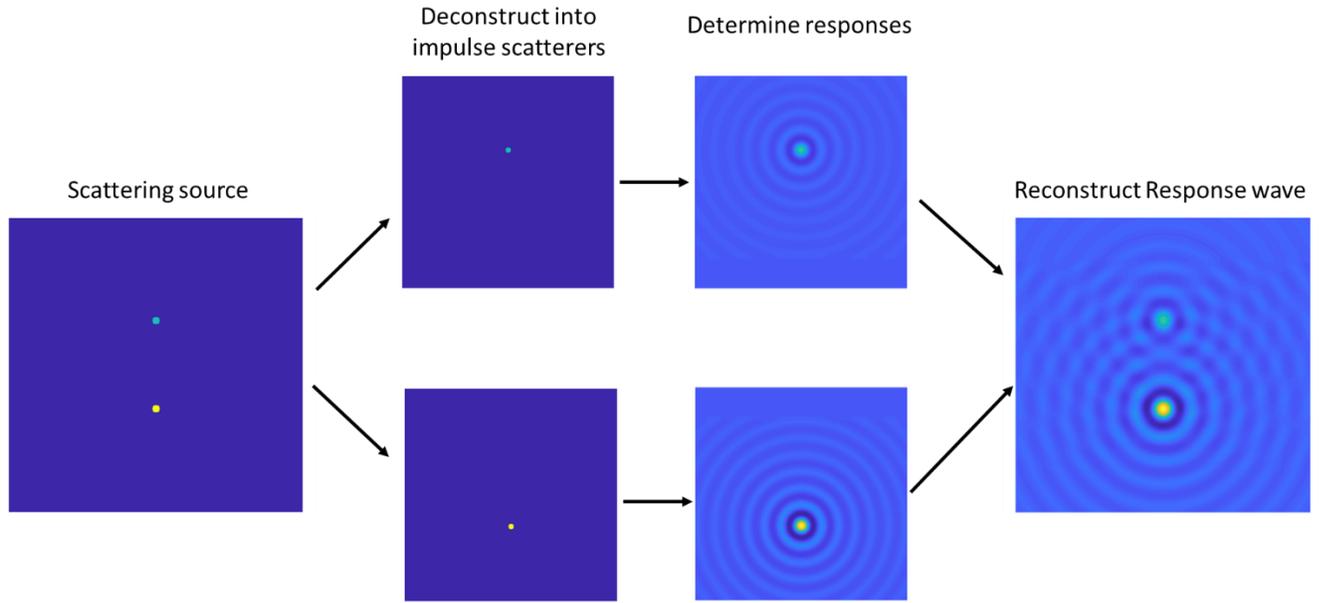

**Figure 6:** Schematic representation of the point source procedure for solving the PDE that describes the scattering field

By exploiting linearity, the solution to Equation (10) can readily be found for any source term $\boldsymbol{Q}$ using the dyadic Green's function $\Gamma(\boldsymbol{r}, \boldsymbol{r}')$, which provides the electric field solution at $\boldsymbol{r}$ to a point source at $\boldsymbol{r}'$. **Figure 6** demonstrates the way in which the solution is constructed. Linearity implies that a sum of sources yields a sum of responses. An arbitrary source is decomposed into a sum of impulse sources, each of which generates its own response (provided by the Green's function). The responses are summed together to construct the solution to the arbitrary source.

Formally, for each point source location $\boldsymbol{r}'$, the Green's function must satisfy

$$\nabla \times \left( \nabla \times \Gamma(\boldsymbol{r}, \boldsymbol{r}') \right) - k_0^2 \bar{\epsilon}\, \Gamma(\boldsymbol{r}, \boldsymbol{r}') = \delta(\boldsymbol{r} - \boldsymbol{r}') \tag{30}$$

Where $\delta(\boldsymbol{r} - \boldsymbol{r}')$ is the 3D Dirac-delta generalized function. The far-field solution to equation (10) is

$$\widetilde{\boldsymbol{E}}(\boldsymbol{r}) = \iiint_V d\boldsymbol{r}'\, \Gamma(\boldsymbol{r}, \boldsymbol{r}')\boldsymbol{Q}(\boldsymbol{r}') \tag{31}$$

Which can be verified by substitution into (10)

$$\nabla \times (\nabla \times \widetilde{E}) - k_0^2 \bar{\epsilon} \widetilde{E} = \iiint_V dr' [\nabla \times (\nabla \times \Gamma(r,r')) - k_0^2 \bar{\epsilon} \Gamma(r,r')] Q(r')$$
$$= \iiint_V dr' \delta(r-r') Q(r') = Q \quad (32)$$

Since sources are orientable, the Green's function is most generally dyadic (that is, a tensor), in which case the product (31) is an inner dot product. The exact dyadic Green's function for any lossless, anisotropic $\bar{\epsilon}$ is known [39], but is needlessly complicated for most applications in dielectric scattering. Instead, media which are weakly anisotropic, may have their dyadic Green's function approximated by an isotropic dyadic Green's function. In the far field, the isotropic dyadic Green's function is

$$\Gamma(r,r') \overset{r \gg 2k}{\approx} \frac{e^{i\bar{n}k_0|r-r'|}}{4\pi|r-r'|}\left(I_d - \frac{(r-r') \otimes (r-r')}{|r-r'|^2}\right) \quad (33)$$

Where $\bar{n}$ is an isotropic index of refraction, $I_d$ is the identity matrix (notated $I_d$ to prevent confusion with intensity $I$), and $\otimes$ is the complex dyadic product (linear in the first argument and antilinear in the second). $\bar{n}$ is defined as

$$\bar{n} := \sqrt{\langle \bar{\epsilon} \rangle_\Omega} = \sqrt{\frac{1}{3}\mathrm{tr}(\bar{\epsilon})} \quad (34)$$

where $\langle \cdot \rangle_\Omega$ is an average along all orientations. The condition of being weakly anisotropic can be expressed using

$$\|\bar{\epsilon} - \langle \bar{\epsilon} \rangle_\Omega\| \ll |\langle \bar{\epsilon} \rangle_\Omega| \quad (35)$$

Equation (31) provides a solution to equation (10) if the Green's function is known. For weak anisotropy as expressed in condition (34), the simpler Green's function in (33) may be used to solve equation (10) in the far-field.

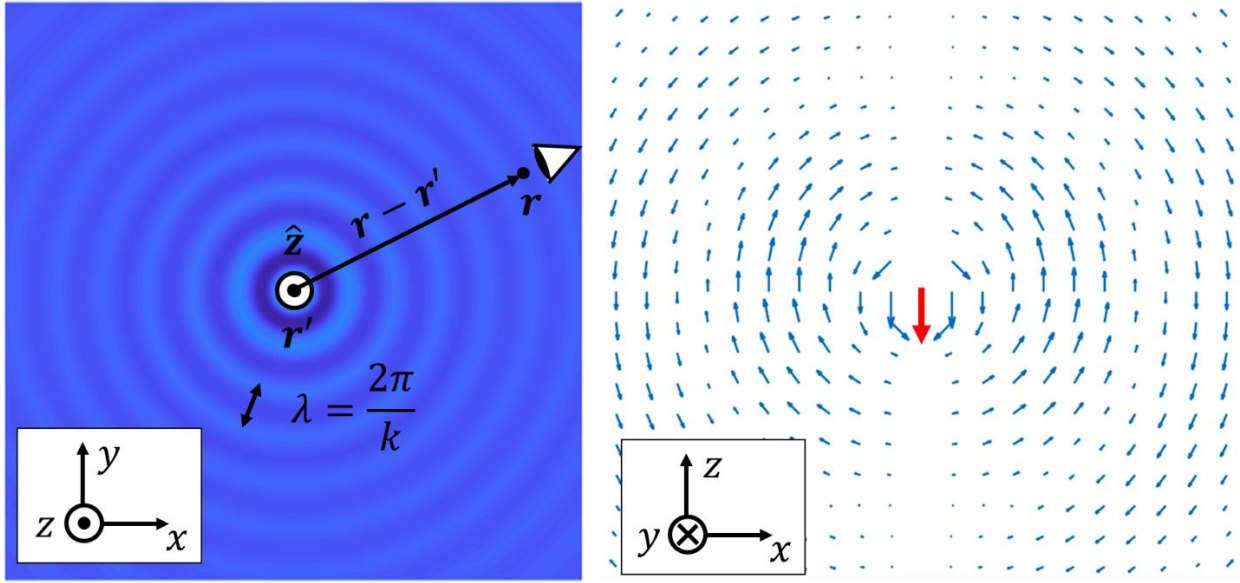

**Figure 7:** The far field dyadic Green's Function for a for a source oscillating along the **z**-direction placed at $r'$. a) z component in the xy-plane and b) xz vector component in the xz-plane

**Figure 7** plots the far field dyadic Green's function as specified in equation (33) for a source placed at $r'$ oscillating along the $z$-direction. The response is symmetric to rotations around the $z$-axis (direction of oscillation). No light is radiated in the $z$-direction, while the light intensity is maximal in the $xy$ plane. This anisotropy in the source scattering field can lead to anisotropy in the overall scattering field if the source term is anisotropic.

### 3. Total scattering intensity

Measurements of scattering by an inhomogeneous sample are typically performed in the far-field, where only propagating modes are relevant. In addition, the far field medium is usually air where the index is $n = 1$. Since the average sample index $\bar{n}$ does not equal 1 in general, there will be reflection and refraction of scattered light as it exits the sample medium. This complicates the exact computation of an angle dependent far-field scattering intensity by requiring the Green's function to account for reflection and refraction at the boundaries of the sample. Since the effect of boundary reflection and refraction depends on the location of the radiation source within the sample, such a Green's function loses space-invariance and become much harder to specify.

However, this problem becomes irrelevant if we restrict our findings to the total amount of energy that is scattered by the medium, regardless of its scattering angle. The key is to note that the total intensity of far-field scattering is not impacted by reflections or refraction of light passing from the sample to the surrounding medium. This is because scattered light energy that is reflected or refracted is still scattered light energy. The reflection and refraction merely serve to redirect the scattered light. Since changing the refractive index of the transparent surrounding medium only affects surface reflection and refraction, the proportion of total light energy scattered remains the same when changing the refractive index. This liberates us to choose any refractive index for the surrounding medium when calculating the total light energy that is scattered. The most convenient surrounding medium to use is one which has index of refraction $n_{ff} \coloneqq \Re(\bar{n})$. This is the real (or Hermitian if a tensor) component of the average sample index $\bar{n}$ as defined in equation (34). This choice of $n_{ff}$ as the index of the medium is one that minimizes reflections at the interface while also being uniform, lossless, and isotropic.

Our goal here will be to calculate the total energy scattered to obtain a scattering coefficient. For this reason, we will solve the scattering problem when the far-field medium is $n_{ff}$, allowing us to ignore the effect of reflection and work with a far-field medium that is ohmic, uniform, lossless, and isotropic. Unless the experiment is performed in a surrounding medium which has index $n_{ff}$, the angle dependent results derived here should not correspond to a measured angle dependent scattering. The inhomogeneous scattering coefficient of the sample as derived here, however, will correspond to a measured scattering coefficient.

### 3.1 Far-field intensity in an ohmic, uniform, weakly anisotropic medium

If the incident light has a coherence length that spans many grains, the incident light and the elastically scattered light can be approximated by monochromatic light. If the far-field medium is also uniform (space-invariant), there is no scattering in the far-field and all sources of scattering originate from the sample. In the far field, the scatterer may be approximated by a point source (which we will place at the origin) emitting spherical waves. Because of this, the scattered, far-field radiation is locally unidirectional (a plane wave) and the Poynting vector $S$ is in the direction $\hat{r}$. If the far-field medium is also weakly anisotropic, then the far-field intensity can be written as

$$I(\boldsymbol{r}, \omega) = \frac{\bar{n}_{ff} \cos \beta}{2\mu_0 c} |\boldsymbol{E}(\boldsymbol{r}, \omega)|^2 = G \cos \beta\, |\boldsymbol{E}(\boldsymbol{r}, \omega)|^2 \tag{36}$$

where $\beta$ is the angle between a collecting surface and the Poynting vector, $c$ is the speed of light in vacuum, $G$ is a proportionality constant (its precise value will turn out to be irrelevant in our final expression of the scattering coefficient), and $\bar{n}_{ff}$ is

$$\bar{n}_{ff} := \sqrt{\langle n_{ff}^2 \rangle_\Omega} = \sqrt{\frac{1}{3} \mathrm{tr}(n_{ff}^2)} \tag{37}$$

A first principles derivation for the validity of equation (36) in weakly anisotropic media can be found in **appendix D**. We will be integrating $I(\boldsymbol{r})$ over the unit sphere in order to compute the total scattering loss. In this case the surface normal will align with $\boldsymbol{r}$ implying $\cos \beta = 1$, in which case

$$I(\boldsymbol{r}) = G|\boldsymbol{E}(\boldsymbol{r}, \omega)|^2 \tag{38}$$

Equation (38) is the same as would be derived for an isotropic far-field medium whose refractive index is $\bar{n}_{ff}$. Thus, there is no drawback in presuming the far field to be isotropic in materials which are weakly anisotropic enough to satisfy (35). Equation (38) is often used as a starting point without much justification. Here we have demonstrated a sufficient set of assumptions that can be used to justify its use.

### 3.2 General expression for far-field scattered intensity

### 3.2.1 From random process

The goal is to compute the ensemble average of the far-field scattering intensity. By choosing $\bar{n}_{ff}$ in the surrounding medium, we have also constructed a far-field medium which is uniform, lossless, and isotropic. In this case, the ensemble average of the far-field scattering intensity is

$$\tilde{I}(\boldsymbol{r}) = G \left\langle |\tilde{\boldsymbol{E}}(\boldsymbol{r})|^2 \right\rangle_W = G \iiint_{\mathcal{V}} d\boldsymbol{r}' \iiint_{\mathcal{V}} d\boldsymbol{r}'' \left\langle \left(\Gamma(\hat{\boldsymbol{r}}, \boldsymbol{r}') \boldsymbol{Q}(\boldsymbol{r}')\right)^\dagger \Gamma(\hat{\boldsymbol{r}}, \boldsymbol{r}'') \boldsymbol{Q}(\boldsymbol{r}'') \right\rangle_W \tag{39}$$

Where $\langle \cdot \rangle_W$ is the average of the ensemble (see **section 2.2**), $\mathcal{V}$ is the set of points in the volume of the scatterer, and the $\dagger$ superscript signifies the adjoint (conjugate transpose). In addition, the mean permittivity $\bar{\epsilon}$ will be averaged over the ensemble so that it will not depend on the (see **section 2.2**). We will assume that the field at $\boldsymbol{r}$ is far enough away so that light measured at $\boldsymbol{r}$ and emitted from a source at $\boldsymbol{r}'$ (near the origin) is approximately unidirectional. By placing the origin within the scattering volume, the following approximations will hold in the far-field

$$r^2 \approx |\boldsymbol{r} - \boldsymbol{r}'||\boldsymbol{r} - \boldsymbol{r}''| \tag{40}$$

$$\hat{\boldsymbol{r}} \approx \frac{\boldsymbol{r} - \boldsymbol{r}'}{|\boldsymbol{r} - \boldsymbol{r}'|} \approx \frac{\boldsymbol{r} - \boldsymbol{r}''}{|\boldsymbol{r} - \boldsymbol{r}''|} \tag{41}$$

$$\begin{aligned} |\boldsymbol{r} - \boldsymbol{r}''| - |\boldsymbol{r} - \boldsymbol{r}'| &= \frac{(\boldsymbol{r} - \boldsymbol{r}'')}{|\boldsymbol{r} - \boldsymbol{r}''|} \cdot (\boldsymbol{r} - \boldsymbol{r}'') - \frac{(\boldsymbol{r} - \boldsymbol{r}')}{|\boldsymbol{r} - \boldsymbol{r}'|} \cdot (\boldsymbol{r} - \boldsymbol{r}') \\ &\approx \hat{\boldsymbol{r}} \cdot (\boldsymbol{r} - \boldsymbol{r}'') - \hat{\boldsymbol{r}} \cdot (\boldsymbol{r} - \boldsymbol{r}') \\ &= \hat{\boldsymbol{r}} \cdot (\boldsymbol{r}' - \boldsymbol{r}'') \end{aligned} \tag{42}$$

Let the function $1_\mathcal{V}(\boldsymbol{r})$ be an indicator function which determines if a point is in the scattering volume.

$$1_\mathcal{V}(\boldsymbol{r}) := \begin{cases} 1 & \boldsymbol{r} \in \mathcal{V} \\ 0 & otherwise \end{cases} \tag{43}$$

Then, since the medium is uniform, the permittivity fluctuation will satisfy

$$\tilde{\epsilon}(r) = 1_{\mathcal{V}}(r)\tilde{\epsilon}(r) \tag{44}$$

Let the incident light be monochromatic, unidirectional with wavevector $k$ and wavenumber $k = \bar{n}k_0$, and linearly polarized in the direction $\widehat{E}_0$ with a complex amplitude $\bar{E}_0$ which is uniform. The incident light is then

$$\bar{E}(r) = \bar{E}_0 \widehat{E}_0 e^{ik \cdot r} \tag{45}$$

Applying this to (11)

$$Q(r) = k_0^2 \bar{E}_0 e^{ik \cdot r} 1_{\mathcal{V}}(r)\tilde{\epsilon}(r)\widehat{E}_0 \tag{46}$$

By introducing the following mediating definitions, we can simplify the resulting equations

$$q := \bar{n}k_0(\hat{r} - \hat{k}) \tag{47}$$

$$\rho := r'' - r' \tag{48}$$

$$\hat{r}_\perp := I_d - \hat{r} \otimes \hat{r} \tag{49}$$

Where $q$ is the momentum transfer in wavenumber units, $\rho$ indexes the separation between two point sources ($r''$ and $r'$) and $\hat{r}_\perp$ is a linear vector operator which takes the perpendicular projection of a vector w.r.t. $\hat{r}$. In the far-field, the approximation

Using equations (33), (39)-(42), (48)-(49), the property $\hat{r}_\perp^H \hat{r}_\perp = \hat{r}_\perp$, and some algebraic manipulation, the scattered intensity is

$$\tilde{I}(r) = \frac{I_0 k_0^4}{(4\pi r)^2} \iiint_\mathcal{V} d\rho \iiint_\mathcal{V} dr' \, 1_{\mathcal{V}}(r')1_{\mathcal{V}}(r' + \rho) \langle \widehat{E}_0^\dagger \tilde{\epsilon}(r')^\dagger \hat{r}_\perp \tilde{\epsilon}(r' + \rho) \widehat{E}_0 \rangle_W e^{-iq \cdot \rho} \tag{50}$$

Where $I_0$ is the intensity of the incident radiation defined as

$$I_0 := G|\bar{E}_0|^2 \tag{51}$$

### 3.2.2 From a space-invariant random process

If the random process is space-invariant, then any average across microstates will be homogeneous. The factor in (50) that is averaged over the ensemble will then be invariant to a shift in $r'$. By selecting $r' = 0$

$$\tilde{I}(r) = I_0 \frac{k_0^4}{(4\pi r)^2} \iiint_\mathcal{V} d\rho \left[ K(\rho, \hat{r}) \iiint_\mathcal{V} dr' \, 1_{\mathcal{V}}(r')1_{\mathcal{V}}(r' + \rho) \right] e^{-iq \cdot \rho} \tag{52}$$

Where $K(\rho, \hat{r})$ is the two-point scattering source autocovariance function. It's role is to quantify the far-field interference of scattering by sources separated by $\rho$. It is defined as

$$K(\rho, \hat{r}) := \langle \widehat{E}_0^\dagger \tilde{\epsilon}(0)^\dagger \hat{r}_\perp \tilde{\epsilon}(\rho) \widehat{E}_0 \rangle_W \tag{53}$$

The outer integral in equation (52) is the Fourier transform of a product of functions. This would correspond to a convolution of the Fourier transforms. The integral inside the brackets is a spatial autocovariance of the indicator function. In Fourier space, this corresponds to the magnitude squared of the Fourier transform. If for an arbitrary function $f(r)$ the Fourier transform is defined as

$$\mathcal{F}\{f(r)\}(q) = \iiint_\mathcal{V} dr \, f(r) e^{-iq \cdot r} \tag{54}$$

with inverse

$$\mathcal{F}^{-1}\{F(\boldsymbol{q})\}(\boldsymbol{r}) = \left(\frac{1}{2\pi}\right)^3 \iiint_V d\boldsymbol{q}\, F(\boldsymbol{q}) e^{i\boldsymbol{q}\cdot\boldsymbol{k}} \tag{55}$$

then equation (52) can then be written as

$$\tilde{I}(\boldsymbol{r}) = I_0 \frac{k_0^4}{(4\pi r)^2}\left[\mathcal{F}_{\boldsymbol{\rho}}\{K(\boldsymbol{\rho},\hat{\boldsymbol{r}})\} \star |\mathcal{F}\{1_v(\boldsymbol{r}')\}|^2\right](\boldsymbol{q}) \tag{56}$$

Where the subscripts of $\mathcal{F}$ and $\mathcal{F}^{-1}$ denote the variable of integration when the input function has more than one argument. $\star$ is the convolution operator defined by

$$[f \star g](\boldsymbol{r}) = \iiint_V d\boldsymbol{r}'\, f(\boldsymbol{r}')g(\boldsymbol{r}-\boldsymbol{r}') \tag{57}$$

A physical interpretation of equation (56) is that space-invariance is broken near the surface of a scattering volume due to the abrupt change in properties at the surface. The convolution describes the effect of the shape of the scattering volume on the scattering intensity. In general, the indicator function can be replaced with an envelope amplitude function to account for gradual changes in the magnitude of the fluctuation.

### 3.3 Bulk approximation

Equation (56) implies that space-invariance can still be a useful approximation if the scattering volume $V$ becomes very large. As $v$ approaches $V$, $|\mathcal{F}\{1_v(\boldsymbol{r}')\}|^2$ diverges, but the average scattering per unit volume does not, instead

$$\lim_{v\to V}\frac{1}{V}|\mathcal{F}\{1_v(\boldsymbol{r}')\}|^2 = \delta(\boldsymbol{\rho}) \tag{58}$$

Where $\delta$ is the Dirac delta function in 3-space. Using (58) in equation (56) is valid when the spatial-frequency bandwidth of the indicator function $1_v$ is much smaller than the spatial-frequency bandwidth of the scattering sources.

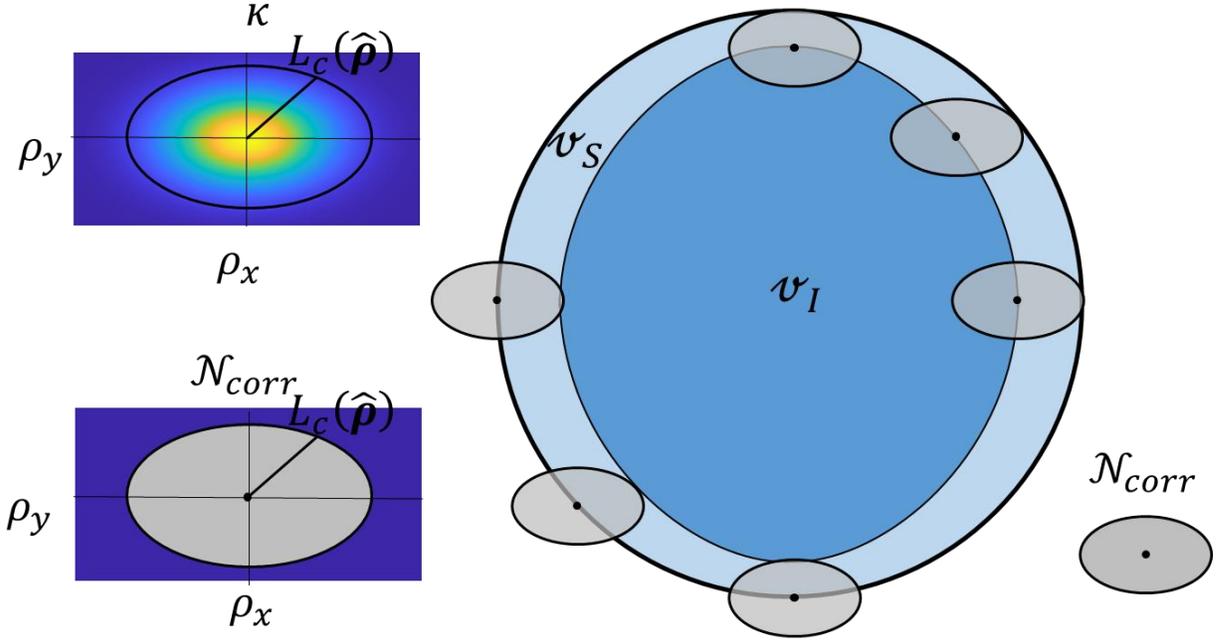

**Figure 8:** Procedure for differentiating exterior (surface) points from interior points.

A physical way to understand this condition is to note that space-invariance can still be a useful approximation if most of the scattering volume is in the interior. **Figure 8** provides a schematic demonstrating the process for discerning interior from exterior points. To do so, a correlation length of the scattering sources $L_c$ can be used (the correlation length is inversely proportional to the spatial-frequency bandwidth of $K$). In general, $L_c$ will be a function of $\hat{\boldsymbol{\rho}}$, the orientation of the shift $\boldsymbol{\rho}$. Points farther than $L_c$ will be assumed to contribute negligibly to the autocovariance function since their scattering contribution $K$ will be nearly zero. Points within $L_c$ are within the correlation neighborhood $\mathcal{N}$ of the central point and cannot be ignored. If all points within a neighborhood are inside the scattering volume, then that point is a member of the interior volume $v_I$. Non-interior points in the scattering volume are members of the surface volume $v_S$ instead. Scattering by interior points are approximated closely by scattering in an infinite scattering volume (since their correlation neighborhood is the same). If the interior volume is much larger than the surface volume then the average scattering per unit volume in $v$ is the same as would occur in an infinite volume $\mathcal{V}$. Formally, the condition

$$|v_I| \gg |v_S| \tag{59}$$

where $|v_I|$ is the volume of interior points, and $|v_S|$ is the volume of surface points allows scattering intensity to be expressed as

$$\tilde{I}(\boldsymbol{r}) = V \left\langle \frac{d\tilde{I}}{dV} \right\rangle_v \approx V \left\langle \frac{d\tilde{I}}{dV} \right\rangle_\mathcal{V} = V \frac{I_0 k_0^4}{(4\pi r)^2} \iiint_\mathcal{V} d\boldsymbol{\rho}\, K(\boldsymbol{\rho}, \hat{\boldsymbol{r}}) e^{-i\boldsymbol{q}\cdot\boldsymbol{\rho}} \tag{60}$$

This is the same equation that can be found by applying the limit in (58) into (56). The scattering intensity in equation (60) is for a bulk sample where surface effects can be neglected.

Since the scattering volume is presumed infinite, the scattering source term is

$$\boldsymbol{Q}(\boldsymbol{r}) = k_0^2 \bar{\boldsymbol{E}}_0 e^{i\boldsymbol{k}\cdot\boldsymbol{r}} \tilde{\epsilon}(\boldsymbol{r}) \hat{\boldsymbol{E}}_0 \tag{61}$$

Comparing this to the definition of $K$ in equation (53), we can express $K$ in terms of the source term

$$K(\boldsymbol{\rho}, \hat{r}) = \frac{G}{k_0^4 I_0} e^{-i\boldsymbol{k}\cdot\boldsymbol{\rho}} \langle \boldsymbol{Q}^\dagger(0) \hat{r}_\perp \boldsymbol{Q}(\boldsymbol{\rho}) \rangle_W \qquad (62)$$

Noting that $\hat{r}_\perp = \hat{r}_\perp^\dagger \hat{r}_\perp$, we may express $K$ in terms of $\boldsymbol{Q}_\perp$ the vector rejection of the source term $\boldsymbol{Q}$ from the observation direction $\hat{r}$ (or, equivalently, the projection of $\boldsymbol{Q}$ onto the plane perpendicular to $\hat{r}$)

$$K(\boldsymbol{\rho}, \hat{r}) = \frac{G}{k_0^4 I_0} e^{-i\boldsymbol{k}\cdot\boldsymbol{\rho}} \langle \boldsymbol{Q}_\perp^\dagger(0) \boldsymbol{Q}_\perp(\boldsymbol{\rho}) \rangle_W \qquad (63)$$

Where it can now be observed that the two-point scattering source autocovariance function $K$ is proportional to the covariance between the projected scattering source at the origin and at the point $\boldsymbol{\rho}$. Substituting equation (63) into (59) gives

$$\tilde{I}(\boldsymbol{r}) = \frac{GV}{(4\pi r)^2} \mathcal{F}\left\{ \langle \boldsymbol{Q}_\perp^\dagger(0) \boldsymbol{Q}_\perp(\boldsymbol{\rho}) \rangle_W \right\}(k\hat{r}) \qquad (64)$$

Where $k\hat{r}$ is the wavevector of the far-field scattered light. Equation (64) makes explicit the mathematical role of the scattering source term. Soft scattering intensity (or more generally, intensity from soft diffraction) is proportional to the Fourier transform of the autocovariance of the projected scattering source field. Using line path ergodicity and heterogeneity in the bulk approximation, we can replace the averaging over $W$ with a volume averaging (see **section 2.2**) yielding

$$\begin{aligned}
\tilde{I}(\boldsymbol{r}) &= \frac{GV}{(4\pi r)^2} \mathcal{F}\left\{ \langle \boldsymbol{Q}_\perp^\dagger(0) \boldsymbol{Q}_\perp(\boldsymbol{\rho}) \rangle_V \right\}(k\hat{r}) \\
&= \frac{G}{(4\pi r)^2} \mathcal{F}\left\{ \iiint_V d\boldsymbol{r}'\, \boldsymbol{Q}_\perp^\dagger(\boldsymbol{r}') \boldsymbol{Q}_\perp(\boldsymbol{r}' + \boldsymbol{\rho}) \right\}(k\hat{r}) \\
&= \frac{G}{(4\pi r)^2} |\mathcal{F}\{\boldsymbol{Q}_\perp\}(k\hat{r})|^2
\end{aligned} \qquad (65)$$

Equation (65) demonstrates that soft scattering intensity at $\boldsymbol{r}$ is proportional to the spatial spectral power density at $k\hat{r}$ of the source term perpendicular to $\hat{r}$. Equation (65) reveals that scattered waves are produced by source waves of the same wavenumber. The projection onto the plane perpendicular to $\hat{r}$ occurs because only transverse electromagnetic waves propagate to the far-field.

### 3.4 Role of strength and spatial coherence of scattering sources

It is sometimes convenient to express the $K$ as a product of the variance $K_0$, which measures the strength of the scattering sources and the normalized autocorrelation $\kappa$, which measures the spatial coherence of the scattering sources. The two are defined as

$$K_0(\hat{r}) \coloneqq K(0, \hat{r}) = \mathrm{Var}_W\left( |\hat{r}_\perp \tilde{\epsilon} \overline{\boldsymbol{E}}_0| \right) = \frac{G}{k_0^4 I_0} \mathrm{Var}_W(Q_\perp) \qquad (66)$$

$$\kappa(\boldsymbol{\rho}, \hat{r}) \coloneqq K(\boldsymbol{\rho}, \hat{r}) / K_0(\hat{r}) \qquad (67)$$

where $\mathrm{var}_W(\cdot)$ is the ensemble variance and $Q_\perp$ is the magnitude of the projection of the scattering source onto the surface perpendicular to the observation direction $\hat{r}$. The advantage is that the scattering source strength $K_0$ is dependent on the strength of the electromagnetic response, but not the spatial distribution of the response. The scattering source autocorrelation $\kappa$, on the other hand, is often dependent on the spatial distribution of the response but not on the response itself. To see this, consider that for the scattering source term $\boldsymbol{Q}$ as expressed in (11) to be correlated, both the permittivity fluctuation $\tilde{\epsilon}$ and the phase of the incident light field $\overline{\boldsymbol{E}}$ need to be correlated. Since in practice the source of permittivity fluctuations (grain structure, thermal noise, defects, etc.) is almost always independent of the correlation of incident light (which is affected by the light source), the overall autocorrelation should be the product of autocorrelations

$$\kappa(\boldsymbol{\rho}, \hat{\boldsymbol{r}}) = \kappa_{\bar{E}}(\boldsymbol{\rho}) \kappa_W(\boldsymbol{\rho}, \hat{\boldsymbol{r}}) \tag{68}$$

Where $\kappa_{\bar{E}}$ is the spatial autocorrelation of the incident light phase and $\kappa_W$ is the structure autocorrelation. $\kappa_{\bar{E}}(\boldsymbol{\rho})$ does not depend on $\hat{\boldsymbol{r}}$ since phase is a scalar field and has no intrinsic orientation. For monochromatic light, $\widehat{\boldsymbol{E}}_0$ is completely correlated everywhere and $\kappa_{\bar{E}}(\boldsymbol{\rho}) = 1$. For non-monochromatic light, a coherence time $\tau_{\bar{E}}$ can be defined from the emission spectrum of the incident light, and from it a coherence length in the material $L_{\bar{E}} = (c/\bar{n})\tau_{\bar{E}}$. In most cases, the correlation length of the incident light is much longer than the structure correlation length ($L_{\bar{E}} \gg L_W$) and it can be assumed that

$$\kappa(\boldsymbol{\rho}, \hat{\boldsymbol{r}}) \approx \kappa_W(\boldsymbol{\rho}, \hat{\boldsymbol{r}}) \tag{69}$$

So that $\kappa$ encodes for the structure only. When the correlation length is much smaller than the structural correlation length, then the structural inhomogeneities are no longer probed by the incident light and, if the first-order perturbation condition (27) is still satisfied, negligible scattering occurs due to inhomogeneities. Substituting equation (67) into (60),

$$\tilde{I}(\boldsymbol{r}) = I_0 V \frac{k_0^4}{(4\pi r)^2} K_0(\hat{\boldsymbol{r}}) \iiint_V d\boldsymbol{\rho}\, \kappa_W(\boldsymbol{\rho}, \hat{\boldsymbol{r}}) e^{-i\boldsymbol{q}\cdot\boldsymbol{\rho}} \tag{70}$$

In equation (70), the strength of the scattering sources $K_0(\hat{\boldsymbol{r}})$ directly determines the strength of the scattering intensity $\tilde{I}(\boldsymbol{r})$. The spatial coherence $\kappa_W(\boldsymbol{\rho}, \hat{\boldsymbol{r}})$ of scattering sources determines, via the integral in (70), determines the effect of far-field interference.

### 3.5 Isotropic random process

If the random process is also isotropic, then no preferred orientation exists when averaging along the ensemble of samples $W$, which allows us to conclude that $\kappa_W$ is a function of $\rho \coloneqq |\boldsymbol{\rho}|$ only. Integration along all orientations of $\boldsymbol{\rho}$ yields

$$\tilde{I}(\boldsymbol{r}) = I_0 V \frac{k_0^4}{4\pi r^2} K_0(\hat{\boldsymbol{r}}) \int_0^\infty d\rho\, \rho^2 \kappa_W(\rho) \text{sinc}(q(\hat{\boldsymbol{r}})\rho) \tag{71}$$

where

$$q(\hat{\boldsymbol{r}}) \coloneqq |\boldsymbol{q}(\hat{\boldsymbol{r}})| = 2\bar{n} k_0 \sin(\varphi/2) \tag{72}$$

and $\varphi$ is the angle between $\boldsymbol{k}$ and $\boldsymbol{r}$.

For space-invariant processes for which most points are in the interior, equation (70) can produce the scattering intensity. If the random process is also isotropic, then equation (71) will also produce the scattering intensity. Equation (71) is simpler due to the triple integral being replaced by a single integral and the microstructural autocorrelation $\kappa_W$ depending only on the shift length $\rho$ and not the direction of the shift. However, methods for finding the scattering source strength $K_0$ and autocorrelation $\kappa_W$ are necessary to take advantage of these equations.

### 3.6. Total scattering

It is possible to find the total scattering $\alpha_{sc}$ by a small scattering volume by integrating the scattering intensity $\tilde{I}$ over all directions $\hat{\boldsymbol{r}}$.

$$\alpha_{sc} = \iint_{\hat{V}} d\hat{\boldsymbol{r}}\, \tilde{I}(\hat{\boldsymbol{r}}) \tag{73}$$

where $\hat{\mathcal{V}}$ is the set of all real valued unit vectors in three dimensions. Using equation (60)

$$\alpha_{sc} = I_0 V \frac{k_0^4}{16\pi^2} \iint_{\hat{\mathcal{V}}} d\hat{r} \iiint_{\mathcal{V}} d\boldsymbol{\rho}\, K(\boldsymbol{\rho}, \hat{r}) e^{-i\boldsymbol{q}(\hat{r})\cdot\boldsymbol{\rho}} \tag{74}$$

For small fluctuations caused by a space-invariant and isotropic random process, equation (71) may be used for the scattering intensity, and by changing the order of integration

$$\alpha_{sc} = I_0 V k_0^4 \int_0^\infty d\rho\, \rho^2 \kappa_W(\rho) \frac{1}{4\pi} \iint_{\hat{\mathcal{V}}} d\hat{r}\, K_0(\hat{r}) \operatorname{sinc}(q(\hat{r})\rho) \tag{75}$$

Equations (74) and (75) provide expressions that solve for the total scattering intensity caused by a small scattering volume. For a large scattering volume, however, the scattering coefficient is more appropriate as a measure of scattering strength.

### 3.7. Scattering coefficient

#### 3.7.1 Full scattering coefficient

For the first-order perturbation to be an accurate representation, the incident field should locally resemble the driving field (the field which drives the scattering sources to oscillate) over length scales comparable to the perturbation coherence length $L_c$. Across long distances, however, the driving field is expected to attenuate due to scattering as it travels through the scattering volume. For unidirectional incident radiation, this behavior can be captured using the Lambert law which can be derived by sectioning the scattering volume into thin slices which are perpendicular to the direction of energy transfer (the pointing vector) of the incident light. If the cross-sectional area of the illuminated portion is $A$ and the thickness of each slice is $d\ell$, then the volume of each slice would be

$$dV = A d\ell \tag{76}$$

Assuming scattered light from inhomogeneities is never recovered (e.g., scattering of each slice is independent; can be invalid if different volume slices have scattering that is correlated), the change in the beam intensity due to scattering losses in each section would be

$$dI = -\frac{d\alpha_{sc}}{A} = -\frac{1}{A}\left(\frac{IdV}{I_0 V}\alpha_{sc}\right) = \left(-\frac{\alpha_{sc}}{I_0 V}\right) I d\ell \tag{77}$$

This equation can be solved to yield the familiar Lambert law

$$I = I_0 \exp(-\mu_{sc}\ell) \tag{78}$$

Where the scattering coefficient is

$$\mu_{sc} = \frac{\alpha_{sc}}{I_0 V} = \frac{k_0^4}{16\pi^2} \iint_{\hat{\mathcal{V}}} d\hat{r} \iiint_{\mathcal{V}} d\boldsymbol{\rho}\, K(\boldsymbol{\rho}, \hat{r}) e^{-i\boldsymbol{q}(\hat{r})\cdot\boldsymbol{\rho}} \tag{79}$$

For isotropic random processes equation (79) becomes

$$\mu_{sc} = k_0^4 \int_0^\infty d\rho\, \rho^2 \kappa_W(\rho) \frac{1}{4\pi} \iint_{\hat{\mathcal{V}}} d\hat{r}\, K_0(\hat{r}) \operatorname{sinc}(q(\hat{r})\rho) \tag{80}$$

The scattering cross sections $\mu_{sc}$ given by (79) and (80) quantify the scattering losses that are due to inhomogeneities in refractive index. The quintuple integral in (79) and the triple integral in (80) must be performed for each value of $k_0$. So, while these results are extremely general, they may be slow to execute when computing $\mu_{sc}$ as a function of $k_0$ or $\lambda$. For this reason, we will investigate approximations to these solutions under more specific conditions.

### 3.7.2 Isotropic random process in absorption-less media

In scattering applications resulting from isotropic random processes in passive, lossless media $K_0(\hat{r})$ can be approximated by a constant proportional to the variance of $\epsilon$

$$K_0(\hat{r}) \approx K' \text{var}_W(\epsilon) \tag{81}$$

where $K'$ is a proportionality constant which takes on a value around 0.7. Being isotropic, passive, and lossless is sufficient to derive (81) but is not necessary. In fact, equation (81) is valid whenever the inhomogeneities in permittivity are approximately Hermitian. This occurs when then inhomogeneities of absorption (the anti-Hermitian component of the refractive index tensor $n$) are much smaller than the inhomogeneities of refractive index. This is often the case when the absorption is weak or non-existent. Even for stronger absorption, if the absorption sites are isotropic, then again the refractive index inhomogeneities are approximately Hermitian. We plan to publish a derivation demonstrating equation (81) for Hermitian refractive indices. In this paper, however, we will only investigate the consequence of (81) being valid. Equation (80) then reduces to

$$\mu_{sc} = K' \frac{k_0^2}{\bar{n}^2} \text{var}_W(\epsilon) \int_0^\infty d\rho \, \kappa_W(\rho) \sin^2(k\rho) \tag{82}$$

When the RGDA condition (23) is satisfied, then the variance of the index $n$ can be approximated by

$$\text{var}_W(n) = \text{var}_W(\sqrt{\epsilon}) = \text{var}_W(\sqrt{\bar{\epsilon} + \tilde{\epsilon}}) \approx \text{var}_W\left(\sqrt{\bar{\epsilon}} + \frac{\tilde{\epsilon}}{2\sqrt{\bar{\epsilon}}}\right) = \frac{\text{var}_W(\tilde{\epsilon})}{4\bar{\epsilon}} = \frac{\text{var}_W(\epsilon)}{4\bar{n}^2} \tag{83}$$

Or equivalently

$$\text{var}_W(\epsilon) = 4\bar{n}^2 \text{var}_W(n) \tag{84}$$

Substituting this into (82)

$$\mu_{sc} = 4K' k_0^2 \text{var}_W(n) \int_0^\infty d\rho \, \kappa_W(\rho) \sin^2(\bar{n} k_0 \rho) \tag{85}$$

Equation (85) is simpler to compute than (80), easier to manipulate, and more readily lends itself to interpretation. Because of this, equation (85) is preferred over equation (80) in isotropic absorption-less media.

### 3.7.3 Rayleigh small-size limit

In reality, the integral over $\rho$ is never infinite. Instead, it is sufficient to integrate up to some multiple of the correlation length $L_c$ of scattering sources. If the correlation length is small enough to satisfy

$$\bar{n} k_0 L_c \ll 1 \tag{86}$$

then $q(\hat{r})\rho \ll 1$ and a first-order approximation of the exponential in equation (79) yields

$$\mu_{sc} = \frac{k_0^4}{16\pi^2} \iint_{\hat{\mathcal{V}}} d\hat{r} \iiint_{\mathcal{V}} d\boldsymbol{\rho} \, K(\boldsymbol{\rho}, \hat{r})(1 - i\boldsymbol{q} \cdot \boldsymbol{\rho}) \tag{87}$$

Exploiting the property inherited from space-invariance

$$K(\boldsymbol{\rho}, \hat{r}) = \langle \widehat{\boldsymbol{E}}_0^\dagger \tilde{\epsilon}(0)^\dagger \hat{r}_\perp \tilde{\epsilon}(\boldsymbol{\rho}) \widehat{\boldsymbol{E}}_0 \rangle_W = \langle \widehat{\boldsymbol{E}}_0^\dagger \tilde{\epsilon}(-\boldsymbol{\rho})^\dagger \hat{r}_\perp \tilde{\epsilon}(0) \widehat{\boldsymbol{E}}_0 \rangle_W \tag{88}$$

$$= \langle \hat{E}_0^\dagger \tilde{\epsilon}(0)^\dagger \hat{r}_\perp \tilde{\epsilon}(-\boldsymbol{\rho}) \hat{E}_0 \rangle_W^* = K(-\boldsymbol{\rho}, \hat{r})^*$$

$$\mu_{sc} = \frac{k_0^4}{16\pi^2} \iint_{\hat{\mathcal{V}}} d\hat{r} \left( \iiint_{\mathcal{V}} d\boldsymbol{\rho}\, K(\boldsymbol{\rho}, \hat{r}) + \iiint_{\mathcal{V}} d\boldsymbol{\rho}\, \Im[K(\boldsymbol{\rho}, \hat{r})](\boldsymbol{q} \cdot \boldsymbol{\rho}) \right) \tag{89}$$

If the medium is non-absorbing (dielectric), then $\epsilon$ is Hermitian and $K$ will have no imaginary component and the second integral vanishes, yielding

$$\mu_{sc} = \frac{k_0^4}{16\pi^2} \iint_{\hat{\mathcal{V}}} d\hat{r} \iiint_{\mathcal{V}} d\boldsymbol{\rho}\, K(\boldsymbol{\rho}, \hat{r}) \tag{90}$$

This is the Rayleigh scattering limit and is valid if the fluctuations are small enough so that condition (86) is satisfied. The advantage of using (90) is that the quintuple integral needs only be performed once, while (79) requires integrating over each value of $k_0$.

### 3.7.3.1 Isotropy

In the isotropic case, $\mathrm{sinc}(q(\hat{r})\rho) \approx 1$ in the Rayleigh limit and equation (80) becomes

$$\mu_{sc} = k_0^4 \langle K_0 \rangle_{\hat{r}} \int_0^\infty d\rho\, \rho^2 \kappa_W(\rho) \tag{91}$$

Where $\langle \cdot \rangle_{\hat{r}}$ is the average along all observation orientations $\hat{r}$. Defined for a function $x$ of $\hat{r}$ as

$$\langle x \rangle_{\hat{r}} := \frac{1}{4\pi} \iint_{\hat{\mathcal{V}}} d\hat{r}\, x(\hat{r}) \tag{92}$$

$\langle K_0 \rangle_{\hat{r}}$ is the scattering strength averaged along all orientations, it is defined as

$$\langle K_0 \rangle_{\hat{r}} := \frac{1}{4\pi} \iint_{\hat{\mathcal{V}}} d\hat{r}\, K_0(\hat{r}) \tag{93}$$

If the medium is also absorption-less, than (85) becomes

$$\mu_{sc} = 4K' k_0^4 \bar{n}^2 \mathrm{var}_W(n) \int_0^\infty d\rho\, \rho^2 \kappa_W(\rho) \tag{94}$$

### 3.7.3.2 When $\kappa_W$ does not depend on $\hat{r}$

One interpretation of autocorrelation is as the linear transformation resulting from least-squares multivariate regression of a signal. Often, the best-fit linear transformation mapping the scattering source at the origin $\boldsymbol{Q}(\boldsymbol{0})$ to the scattering source at another point $\boldsymbol{Q}(\boldsymbol{\rho})$ is well approximated by a scaling transformation. When this is the case, $\kappa_W$ loses its $\hat{r}$ dependence and becomes a function of $\boldsymbol{\rho}$ only expressed as

$$\kappa_W(\boldsymbol{\rho}, \hat{r}) \approx \kappa_W(\boldsymbol{\rho}) \tag{95}$$

Assuming (95) to be the case leads to a Rayleigh limit expression for $\mu_{sc}$ with a clearer interpretation:

$$\mu_{sc} = \frac{k_0^4}{16\pi^2} \iint_{\hat{\mathcal{V}}} d\hat{r}\, K_0(\hat{r}) \iiint_{\mathcal{V}} d\boldsymbol{\rho}\, \kappa_W(\boldsymbol{\rho}) = \frac{k_0^4 V_{corr}}{4\pi} \left( \frac{1}{4\pi} \iint_{\hat{\mathcal{V}}} d\hat{r}\, K_0(\hat{r}) \right) \tag{96}$$

$$= \frac{k_0^4 \langle K_0 \rangle_{\hat{r}} V_{corr}}{4\pi}$$

where $V_{corr}$ is the correlation volume, a quantity which expresses the volume that, on average, correlates with a point. It is defined as

$$V_{corr} := \iiint_V d\boldsymbol{\rho}\, \kappa_W(\boldsymbol{\rho}) \tag{97}$$

Equations (96) and (97) provide a convenient way to calculate the scattering coefficient when $\kappa_W$ does not depend on $\hat{r}$. We can cast equation (96) in a less concrete, but more meaningful way. To do so, re-express $\langle K_0 \rangle_{\hat{r}}$ as

$$\langle K_0 \rangle_{\hat{r}} = \langle \widehat{\boldsymbol{E}}_0^\dagger \tilde{\epsilon}(0)^\dagger \langle \hat{r}_\perp \rangle_{\hat{r}} \tilde{\epsilon}(0) \widehat{\boldsymbol{E}}_0 \rangle_W = \frac{2}{3} \langle \widehat{\boldsymbol{E}}_0^\dagger \tilde{\epsilon}(0)^\dagger \tilde{\epsilon}(0) \widehat{\boldsymbol{E}}_0 \rangle_W = \frac{2}{3} \mathrm{Var}_W(|\tilde{\epsilon}\widehat{\boldsymbol{E}}_0|)$$
$$= \frac{2G}{3k_0^4 I_0} \mathrm{Var}_W(Q) \tag{98}$$

where $\mathrm{Var}_W(Q)$ is the variance of the scattering source amplitude. Substituting this result

$$\mu_{sc} = \frac{G}{6\pi I_0} V_{corr} \mathrm{Var}_W(Q) \tag{99}$$

We see that the scattering coefficient is proportional to the correlation volume and the variance of the scattering source amplitude.

### 3.7.4. Large-size RGD limit

#### 3.7.4.1 Isotropy

For isotropic microstructures, equation (85) can be written as

$$\mu_{sc} = 2K' k_0^2 \mathrm{var}_W(n) \left( \int_0^\infty d\rho\, \kappa_W(\rho) + \int_0^\infty d\rho\, \kappa_W(\rho) \cos(2\bar{n}k_0\rho) \right) \tag{100}$$

In the large-size limit

$$\bar{n} k_0 L_c \gg 1 \tag{101}$$

and the second integral in (100) vanishes, yielding

$$\mu_{sc} = 2K' k_0^2 \mathrm{var}_W(n) \int_0^\infty d\rho\, \kappa_W(\rho) \tag{102}$$

Which is the large size RGD limit of equation (80). The advantage of using the large size RGD limit is that only one integral needs to be evaluated for the entire range of incident light satisfying (101).

#### 3.7.4.2 When $\kappa_W$ does not depend on $\hat{r}$

Assuming (95) is valid and combining the definition in (67) and equation (79)

$$\mu_{sc} = \frac{k_0^4}{16\pi^2} \iint_{\hat{V}} d\hat{r}\, K_0(\hat{r}) \iiint_V d\boldsymbol{\rho}\, \kappa_W(\boldsymbol{\rho}) e^{-i\boldsymbol{q}(\hat{r})\cdot\boldsymbol{\rho}} = \frac{k_0^4}{16\pi^2} \iint_{\hat{V}} d\hat{r}\, K_0(\hat{r}) \mathcal{F}\{\kappa_W\}(\boldsymbol{q}(\hat{r})) \tag{103}$$

$q(\hat{r})$ traces out a sphere with radius $\bar{n}k_0$ centered at $\boldsymbol{k} = k\hat{\boldsymbol{k}} = \bar{n}k_0\hat{\boldsymbol{k}}$. This is the Ewald sphere and represents all waves in the material relevant to elastic scattering. In the large size limit

$$\bar{n}k_0 L_\perp \gg 1 \tag{104}$$

Where $L_\perp$ is the correlation length perpendicular to the direction of propagation $\hat{\boldsymbol{k}}$. When (104) is satisfied, the Ewald sphere can be approximated by the plane perpendicular to $\hat{\boldsymbol{k}}$. **Figure 9** schematically illustrates the integral required in equation (103) for the small-size, intermediate-size, and large size regimes.

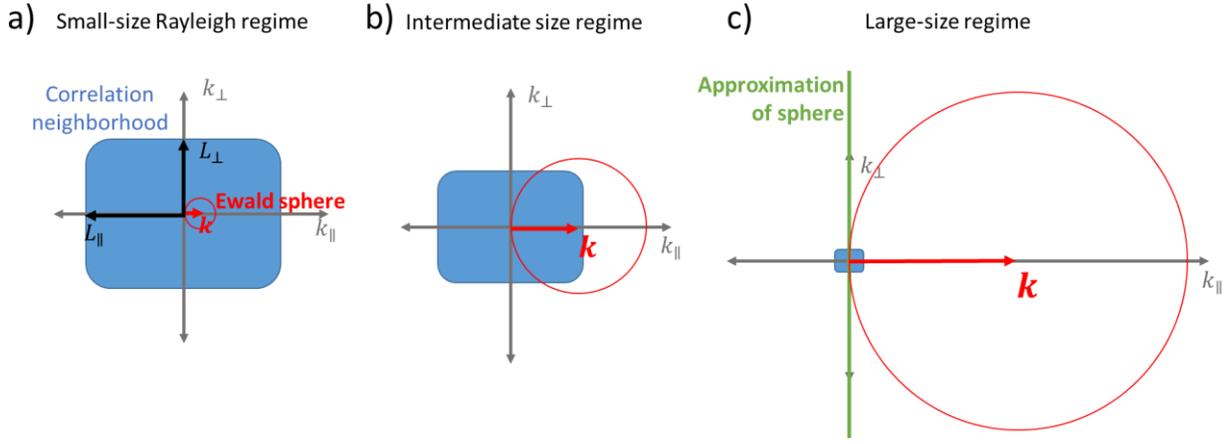

**Figure 9:** illustration of scattering calculation in Fourier space in different size regimes. The integral is of the scattering source correlation $\kappa$ and is taken in the Fourier space along the Ewald sphere. The correlation neighborhood indicates the region where the magnitude $|\mathcal{F}\{\kappa\}|$ is too large to be neglected.

In the small-size regime, **figure 9a** illustrates that the Ewald sphere is small enough that it is nearly zero everywhere so that $\boldsymbol{q} \approx 0$ is a valid approximation, this is an alternate way of deriving (96). In the large-size regime, **figure 9c** illustrates that the Ewald sphere becomes large enough that it can be approximated by a plane. The plane approximation is accurate in the neighborhood where $\hat{\boldsymbol{r}} \approx \hat{\boldsymbol{k}}$. The result is that equation (103) can be expressed as

$$\mu_{sc} \approx \frac{k_0^4 K_0(\hat{\boldsymbol{k}})}{16\pi^2 (\bar{n}k_0)^2} \iint_{\mathcal{V} \perp \hat{\boldsymbol{k}}} d\boldsymbol{q}\, \mathcal{F}\{\kappa_W\}(\boldsymbol{q}) \tag{105}$$

Where the factor $(\bar{n}k_0)^2$ is introduced to account for the scaling of the plane by the square of the radius and $\mathcal{V} \perp \hat{\boldsymbol{k}}$ is all vectors perpendicular to $\hat{\boldsymbol{k}}$. $\mathcal{V} \perp \hat{\boldsymbol{k}}$ can also be thought of as the plane perpendicular to $\hat{\boldsymbol{k}}$ which intersects the origin. Equation (105) can be written as an inner product

$$\mu_{sc} = \frac{k_0^4 K_0(\hat{\boldsymbol{k}})}{16\pi^2 (\bar{n}k_0)^2} \iiint_{\mathcal{V}} d\boldsymbol{q}\, \delta_k(\boldsymbol{q})\, \mathcal{F}\{\kappa_W\}(\boldsymbol{q}) = \frac{k_0^2 K_0(\hat{\boldsymbol{k}})}{16\pi^2 \bar{n}^2} \langle \delta_k, \mathcal{F}\{\kappa_W\}\rangle \tag{106}$$

Where $\delta_k(\boldsymbol{q})$ is a Dirac surface delta function defined as

$$\delta_k(\boldsymbol{q}) = \delta(\boldsymbol{q} \cdot \widehat{\boldsymbol{k}}) \tag{107}$$

Where $\delta$ is the one-dimensional Dirac delta function. The inner product in equation (106) is defined as

$$\langle f, g \rangle = \iiint_V d\boldsymbol{q}\, f^*(\boldsymbol{q}) g(\boldsymbol{q}) \tag{108}$$

Where the superscript $*$ denotes complex conjugate. The 3D Fourier transform (and it's inverse) preserve the inner product up to a $(2\pi)^3$ scaling and (106) can be re-expressed as

$$\mu_{sc} = \frac{k_0^2 K_0(\widehat{\boldsymbol{k}})}{16\pi^2 \bar{n}^2} (2\pi)^3 \langle \mathcal{F}^{-1}\{\delta_k\}, \kappa_W \rangle \tag{109}$$

The inverse Fourier transform of $\delta_k$ can be evaluated directly using an axis in which $\widehat{\boldsymbol{z}} = \widehat{\boldsymbol{k}}$

$$\begin{aligned}
\mathcal{F}^{-1}\{\delta_k\}(\boldsymbol{r}) &= \frac{1}{(2\pi)^3} \iiint_V d\boldsymbol{q}\, \delta_k(\boldsymbol{q}) e^{i\boldsymbol{q}\cdot\boldsymbol{r}} \\
&= \frac{1}{(2\pi)^3} \int_{-\infty}^{\infty} dq_x \int_{-\infty}^{\infty} dq_y \int_{-\infty}^{\infty} dq_z\, \delta(q_z) e^{i(q_x x + q_y y + q_z z)} \\
&= \frac{1}{(2\pi)^3} \left( \int_{-\infty}^{\infty} dq_x\, e^{iq_x x} \right) \left( \int_{-\infty}^{\infty} dq_y\, e^{iq_y y} \right) \left( \int_{-\infty}^{\infty} dq_z\, \delta(q_z) e^{iq_z z} \right) \\
&= \frac{1}{2\pi} \delta(x)\delta(y)
\end{aligned} \tag{110}$$

Substituting into equation (109)

$$\begin{aligned}
\mu_{sc} &= \frac{k_0^2 K_0(\widehat{\boldsymbol{k}})}{4\bar{n}^2} \langle \delta(x)\delta(y), \kappa_W \rangle = \frac{k_0^2 K_0(\widehat{\boldsymbol{k}})}{4\bar{n}^2} \int_{-\infty}^{\infty} dk\, \kappa_W(k) = \frac{k_0^2 K_0(\widehat{\boldsymbol{k}})}{2\bar{n}^2} \int_0^{\infty} dk\, \kappa_W(k) \\
&= \frac{k_0^2 K_0(\widehat{\boldsymbol{k}})}{2\bar{n}^2} L_{corr}(\widehat{\boldsymbol{k}})
\end{aligned} \tag{111}$$

Where $L_{corr}(\widehat{\boldsymbol{k}})$ is a correlation length in the propagation direction, and its value expresses the length that, on average, correlates with a point in the direction $\widehat{\boldsymbol{k}}$. Using RGDA condition (23), it is possible to show that

$$K_0(\widehat{\boldsymbol{k}}) \approx 4\bar{n}^2 \text{Var}_W(|\hat{k}_\perp n \widehat{\boldsymbol{E}}_0|) \tag{112}$$

Where $\hat{k}_\perp$ is an operator that returns the vector rejection onto $\widehat{\boldsymbol{k}}$ and is defined analogous to (49) as

$$\hat{k}_\perp := I_d - \widehat{\boldsymbol{k}} \otimes \widehat{\boldsymbol{k}} \tag{113}$$

Substituting (112) into (111)

$$\mu_{sc} = 2L_{corr}(\widehat{\boldsymbol{k}}) \text{Var}_W(|k_0 \hat{k}_\perp n \widehat{\boldsymbol{E}}_0|) = 2L_{corr}(\widehat{\boldsymbol{k}}) \text{Var}_W(|\boldsymbol{k} \times \widehat{\boldsymbol{E}}_0|) \tag{114}$$

The term $|\boldsymbol{k} \times \widehat{\boldsymbol{E}}_0|$ can be interpreted as a phase delay coefficient experienced by light travelling in the $\widehat{\boldsymbol{k}}$ direction (note that if $n$ is scalar, the term reduces to $k_0 n = \frac{d\phi}{d\ell}$). With this in mind, we can write

$$\mu_{sc} = 2L_{corr}(\widehat{\boldsymbol{k}}) \text{Var}_W\left(\frac{d\phi_\perp}{dl}\right) \tag{115}$$

Equation (115) demonstrates that in the large-size limit, the scattering coefficient is proportional to the correlation length in the direction of the incident beam and the variance in phase accumulation.

**4. Discussion**

**TABLE 2:** List of assumptions

| # | Name | Assumption |
|---|---|---|
| <9> | Non-magnetic medium | Magnetic response of medium is negligible making $\mu = 1$ a valid approx. |
| <10> | Passive medium | A medium which does not produce more radiation than is incident |
| <11> | Lossless medium | A medium which does not absorb radiation |
| <12> | First-order perturbation approximation; Rayleigh-Gans-Debye approximation (RGDA) | Inhomogeneous scattering can be approximated by a first order perturbation of the permittivity field |
| <13> | Soft scattering | Strength of permittivity fluctuations are much smaller than the average permittivity. See equation (23) |
| <14> | Small phase difference | Points with correlated permittivity fluctuations experience approximately the same phase delay in the full solution as they do in the incident solution. See equation (27) |
| <15> | Weak anisotropy | Anisotropy is weak enough that second order effects on scattering may be ignored. See equation (31) |
| <16> | Far-field | Measurements occur far from the source of radiation; radiation source is approximately point source |
| <17> | Ideal far-field medium | Far field is weakly isotropic, uniform, passive, lossless, and minimizes reflections |
| <18> | Uniform | Space invariant |
| <19> | Unidirectional | Solution for which there is only one wavevector $\boldsymbol{k}$ |
| <20> | Isotropic; unaligned polycrystal | Rotation invariant; rotation invariant *statistics* when used to describe a random process |
| <21> | Ensemble ergodicity | There exists an ensemble averaged scattering intensity and it is approached by the volume averaged scattering intensity of a sample for an arbitrarily large volume (not to be confused with Markov ergodicity) |
| <22> | Linearly polarized | Solution for which there is only one electric field direction $\widehat{\boldsymbol{E}}_0$ |
| <23> | Ideal incident radiation | Incident radiation is unidirectional. Requires perfectly monochromatic light. Approximately valid for incident light when source is far away and the correlation length of incident light amplitude is much larger than the structure correlation length |
| <24> | Homogeneous process | Having uniform statistics. Approximately valid if medium is homogeneous in the interior and is interior volume dominated |
| <25> | Interior volume dominated | Interior volume is much larger than surface volume |
| <28> | Dielectric | Non-conducting |
| <29> | Independent scattering | Interference between scattering sources does not occur; intensity of scattering is the sum of intensities of the sources; In Beer-Lambert law, |
| | | attenuation of incident radiation occurs in length scales much larger than the correlation length of sources $L_c$ ($\mu_{sc} L_c \ll 1$) |
| <30> | Isotropic sources | Sources scatter uniformly in each direction |
| <31> | $\hat{r}$-independent autocorrelation | Source autocorrelation $\kappa$ does not depend on the observation direction $\hat{r}$ |
| <32> | Rayleigh approximation; Rayleigh limit | Approximate scattering valid for dielectrics when their size parameter is much less than 1 |
| <33> | Large-size RGD limit | RGDA when the size parameter is much greater than 1 |

As mentioned earlier, Some of the equations derived in this paper have been derived before. The closest to this work was the Debye and Beuche [13], Goldstein and Michalik [14], Ross and Javis [16-17] line of research. Goldstein [14] and Ross [17] also derive scattering equations for inhomogeneities that are anisotropic. Our work improves on the existing derivations by making them more clear, general, and meaningful.

In the interest of clarity, we disclose analysis which is often taken for granted, such as the use of a stochastic ensemble (**section 2.2**), the assumptions required of the far-field medium (**section 3.1**), and the assumptions required of the scattering volume (**section 3.3**). We have also provided clear definitions for all variables and operations. For instance, averaging requires a sample space with a probability measure and the often used $\langle \cdot \rangle$ leaves the underlying probability measure opaque. Care has been taken to tag the different averages (e.g. ensemble average $\langle \cdot \rangle_W$ vs volume average $\langle \cdot \rangle_V$) in order to be more clear about the expectation operation in question.

By introducing assumptions strategically, the results are organized to maximize their generality. For example, results are first given for soft scattering in general (**section 3.2.1**), then for a process that is space-invariant (**section 3.2.2**), then for a process that is also bulk (**section 3.3**), and then for a process that is also isotropic (**section 3.5**). This allows future researchers to select the level of generality that is appropriate for them. Some results have also been generalized. In the first paper in the series [1], it was shown that constructing an orthogonal principal axis is permitted even for non-orthorhombic crystal systems by showing that lossless media, regardless of crystal system, are always diagonalizable. In addition, **section 3.7.4** generalizes the large-size RGD scattering equations for anisotropic fluctuations.

The results have been made more meaningful by applying them to the calculation of scattering coefficients, which can be extracted from in-line transmission measurements [25]. We generalized the application of our model by arguing (**section 3**) that changing the far-field medium does not affect the scattering coefficient, making the modeling of scattering coefficient more widely applicable than the scattering intensity provided in other works. Wherever possible, care was taken to make the equations more meaningful by providing physical interpretations of variables used and the relationships between them. For example, equation (65) is derived in order to elucidate the relationship between excess accelerating charges and the intensity of soft scattered light.

In the first paper in this series [1], we derived an elastic scattering equation from Maxwell's equations. In this paper, we provide a first principles account of the work of much of the RGD literature by perturbing the scattering equation to evaluate soft scattering in dielectrics due to inhomogeneities. Equations for the scattered electric field and the total scattering intensity were derived for the first-order perturbation (whose validity was studied in **section 2.3**), culminating in a derivation of the scattering coefficient in equation (79). Simpler expressions for the scattering coefficient can be found in equation (80) for isotropic random processes and in equation (85) absorption-less media. Table 2 summarizes the assumptions made at different points in the analysis.

Further simplifications may be made for the small-size (**section 3.7.3**) and large-size (**section 3.7.4**) regimes. **Figure 10** summarizes the expressions for the scattering coefficients in the different regimes for absorption-less media. It also visually demonstrates the effect of modulating by $\sin^2$ before integrating in the different regimes.

A few novel results were derived. In equations (79) (80) and (85), different versions of the scattering intensity are integrated to produce equations which connect the inhomogeneities to the scattering coefficient. The model predictions can then be compared with measurements of the scattering coefficient, which can be extracted from in-line transmission measurements. In equation (56), an equation that describes the effect of scattering volume geometry on the scattering intensity is provided. Furthermore, equation (111) provides an equation for the large-size RGD scattering coefficient. The equation is, fittingly, reminiscent of the eikonal approach generalized to account for a non-scalar refractive index.

Our first principles derivation is instrumental in providing a deeper understanding of the source of inhomogeneous scattering. Physically, the source term in (11) is proportional to the excess (or, if negative, lack of) accelerating charge. "Excess" refers to acceleration larger than would be produced in a uniform

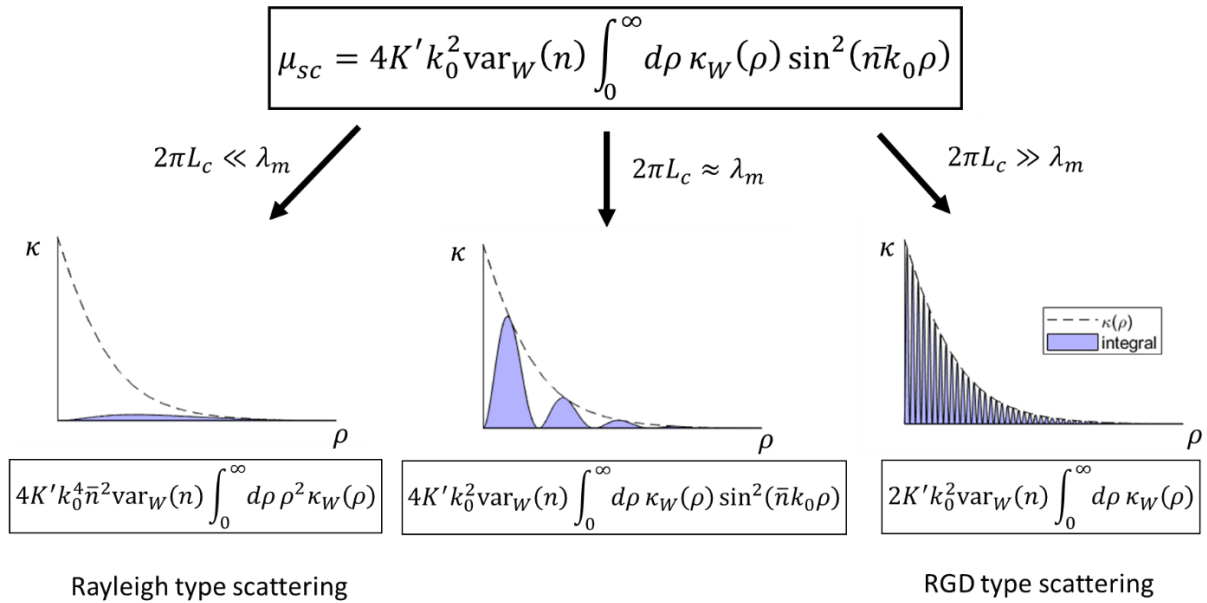

**Figure 10:** Summary of the expressions that are appropriate in different size regimes. The dashed lines are scattering source autocorrelation $\kappa(\rho)$ and the blue shaded regions are the integrals in the equations. The integral in the large size regime is approximately half of the total area under $\kappa(\rho)$

sample with average permittivity. Accelerating charges will produce electromagnetic radiation and behave as a radiation source. In a uniform transparent medium, radiation sources constructively interfere to produce an apparent, slower wave in the direction of propagation while destructively interfering in every other direction. However, the excess scattering sources, which are proportional to the source term $Q$, will not be appropriately compensated for, leading to imperfect constructive interference in the propagation direction (loss of in-line transmission), and imperfect destructive interference in other directions (scattering). We see then that the sources of scattering in inhomogeneous media are the excess accelerating charges. This contrasts with the frequent claim [26-36] that boundaries scatter in polycrystalline transparent ceramics. In almost all cases, grain boundaries themselves do not scatter, rather it is the deviation of the optical properties from the mean that scatters by leading to excess accelerating charges. The exceptions to this are when the RGDA assumptions are not met, or when a significant volume fraction of the medium is in or near boundary regions, in which case grain boundaries can be thought of as a secondary phase that scatters. The mathematical role of scattering sources on the far field scattering intensity was revealed by equation (65), where it can be seen that the far-field scattering intensity at $r$ is proportional to the source term power density at wavevector $k\hat{r}$.

Understanding that RGD theory is a first order perturbation theory illuminates some important insights. If the average permittivity $\bar{\epsilon}$ is a tensor, then double refraction may still occur in the incident beam. This implies that a global alignment of anisotropy can lead to double refraction in inhomogeneous media. Partial alignment of anisotropic polycrystals, for example, should introduce a weakened birefringence on the macro scale. This phenomena was not predicted in other models which discuss alignment in polycrystals even when using an RGD formulation [14, 31, 40-44]. Another insight that is often overlooked is the origin of the soft-scattering condition (28) [13, 27, 29, 45-51]. The soft-scattering condition is sometimes justified on the grounds that reflections can be ignored [48, 49, 51]. This statement is valid when the particles are large enough to produce reflections since reflections account for the largest effect of the correction terms on light propagation. However, for particles smaller than the wavelength of light the line between reflection and scattering becomes blurred, but the soft scattering condition (28) was shown to endure via the more general construction of condition (23). These arguments required that the incident field approximate the actual field on the scale of the coherence length of inhomogeneities. We can then understand the soft scattering condition more fundamentally as imposing limits on the reduction of the electric field amplitude due to inhomogeneities, whether from reflection, scattering, or another loss mechanism.

The derivation of the scattering coefficient also elucidates overlooked concepts. Equation (38) states that the intensity is proportional to the square of the electric field magnitude and is often used as a starting point without much justification [40, 42, 51-54]. Here we have demonstrated a sufficient set of assumptions (ohmic, uniform, weakly anisotropic medium) that can be used to justify its use. Equation (56) gives an expression for scattering intensity in a sample which is finite. A physical interpretation of equation (56) is that the space-invariance is broken near the surface of a scattering volume due to the abrupt change in properties at the surface. The convolution describes the effect of that inhomogeneity on the scattering intensity. In general, the indicator function can be replaced with an envelope amplitude function to account for changes in the magnitude of the fluctuation. Criterion (59) is a condition for the validity of the bulk approximation (60) which allows one to ignore the effects of the scattering volume surface.

In agreement with Debye and Beuche [13], In **section 3.4** we factored the scattering source autocovariance $K$ into the scattering source strength $K_0$ and the scattering source autocorrelation $\kappa$. The advantage is that the scattering source strength $K_0$ is often dependent on the material properties, but not the spatial distribution. The scattering source autocorrelation $\kappa$, on the other hand, is often dependent on the spatial distribution of the permittivity but not on the value of the permittivity itself. This allows us to find the autocorrelation by investigating two simpler quantities $K_0$ and $\kappa$.

In the case when the scattering source autocorrelation $\kappa$ does not vary with $\hat{r}$, the integrals in equation (79) become separable. This leads to a factoring of the scattering coefficient into two factors. The first quantifying the typical strength of a scattering source and the second quantifying the typical amount of scattering sources that correlate (and thus constructively interfere) with a scattering source. When looking at the Rayleigh regime this results in the scattering coefficient being proportional to the variance of the scattering source magnitude and the correlation volume. In the large-size RGD regime the scattering coefficient is proportional to the variance of phase accumulation and the correlation length.

## 5. Conclusion

In summary we present a general, approachable, and comprehensive rederivation of the RGD scattering theory. Alongside the first paper in the series, we have demonstrated how to derive the inhomogeneous scattering coefficient from Maxwell's microscopic equations. This has revealed a deeper understanding of the assumptions employed whenever an RGD approach is used and indicated potential generalizations of the approach. In particular, the RGD assumptions are understood to be derived from the condition that the incident and resulting radiation within a correlation volume be close, and that the intensity as amplitude-squared relationship is understood to require an ohmic, uniform, and isotropic (or weakly anisotropic) medium. In addition, soft scattering sources are shown to be excess accelerating charges. The RGD scattering theory was also generalized to describe the effect of the geometry of the scattering volume and to predict the effect of double refraction when the permittivity exhibits a global alignment. A list of major equations along with the assumptions required for employing them are given in **Table 3.**

**TABLE 3:** List of Major equations and required assumptions. Equations are denoted using () and assumptions using <>.

| Equation | Conclusion | Eq # | Requisites |
|---|---|---|---|
| $\nabla \times (\nabla \times \widetilde{\boldsymbol{E}}) - k_0^2 \bar{\epsilon}\widetilde{\boldsymbol{E}} = \boldsymbol{Q}$ | PDE for first-order scattering | (10) | (1) <13,14>→<12> |
| $\widetilde{\boldsymbol{E}}(\boldsymbol{r}) = \iiint_V d\boldsymbol{r}' \, \Gamma(\boldsymbol{r},\boldsymbol{r}')\boldsymbol{Q}(\boldsymbol{r}')$ | Solution to first-order scattering | (27) | (10) <15> |
| $\tilde{I}(\boldsymbol{r}) = \dfrac{I_0 k_0^4}{(4\pi r)^2}[\mathcal{F}_\rho\{K(\hat{\boldsymbol{r}})\} \star |\mathcal{F}\{1_v\}|^2](\boldsymbol{q})$ | Describes the effect of sample boundaries on the intensity | (56) | (27,39) <23, 24> |
| $\tilde{I}(\boldsymbol{r}) = \dfrac{G}{(4\pi r)^2}|\mathcal{F}\{\boldsymbol{Q}_\perp\}(\boldsymbol{q})|^2$ | Scattering field as spectral power density | (65) | (56) <25> |
| $\mu_{sc} = \dfrac{1}{I_0 V}\iint_{\hat{v}} d\hat{\boldsymbol{r}}\, \tilde{I}(\hat{\boldsymbol{r}})$ | Scattering coefficient from scattering intensity | (73)+(79) | <29> |
| $\mu_{sc} = \dfrac{k_0^4}{16\pi^2}\iint_{\hat{v}} d\hat{\boldsymbol{r}} \iiint_V d\boldsymbol{\rho}\, K(\boldsymbol{\rho},\hat{\boldsymbol{r}}) e^{-i\boldsymbol{q}(\hat{\boldsymbol{r}})\cdot\boldsymbol{\rho}}$ | Scattering coefficient from soft, inhomogeneous scattering | (79) | (56) <25, 29> |
| $\mu_{sc} = \dfrac{k_0^4}{4\pi}\int_0^\infty d\rho\, \rho^2 \kappa_W(\rho) \iint_{\hat{v}} d\hat{r} K_0(\hat{r}) \mathrm{sinc}(q(\hat{r})\rho)$ | Isotropic scattering coefficient | (80) | (79) <20> |
| $\mu_{sc} = \dfrac{G}{6\pi I_0} V_{corr} \mathrm{Var}_W(Q)$ | Small-size Rayleigh regime | (99) | (79) <31, 32> |
| $\mu_{sc} = 2L_{corr}(\hat{\boldsymbol{k}}) \mathrm{Var}_W\left(\dfrac{d\phi_\perp}{dl}\right)$ | Large-size regime | (115) | (79) <31, 33> |

## 6. Acknowledgements

Support of this work by the Office of Naval Research (ONR) through an MRI form the DE-JTO is most gratefully acknowledged

7. Appendices

**Appendix A: derivation of scattering equation**

We start by assuming that Maxwell's equations can fully describe inhomogeneous elastic scattering and work to solve the equations in the presence of such fluctuations. Maxwell's microscopic equations in differential form are

$$\nabla \cdot \boldsymbol{e} = \rho_e \tag{A-116}$$

$$\nabla \cdot \boldsymbol{b} = 0 \tag{A-117}$$

$$\nabla \times \boldsymbol{e} = -\frac{\partial \boldsymbol{b}}{\partial t} \tag{A-118}$$

$$\nabla \times \boldsymbol{b} = \mu_0 \boldsymbol{j} + \mu_0 \epsilon_0 \frac{\partial \boldsymbol{e}}{\partial t} \tag{A-119}$$

where $\boldsymbol{e}$ is the electric field, $\rho_e$ is the electric charge density, $\boldsymbol{b}$ is the magnetic flux density, $t$ is time, $\boldsymbol{j}$ is the electric current density, $\mu_0$ is the vacuum permeability, and $\epsilon_0$ is the vacuum permittivity. Taking the curl of equation (A-3) and then substituting using equation (A-4)

$$\nabla \times (\nabla \times \boldsymbol{e}) + \epsilon_0 \mu_0 \frac{\partial^2 \boldsymbol{e}}{\partial t^2} = -\mu_0 \frac{\partial \boldsymbol{j}}{\partial t} \tag{A-120}$$

Let us re-express equation (A-5) in operator form

$$[(\nabla \times)^2 + \epsilon_0 \mu_0 \partial_t^2]\boldsymbol{e} = -\mu_0 \partial_t \boldsymbol{j} \tag{A-121}$$

Where $\nabla \times$ denotes the curl operator, $(\nabla \times)^2$ denotes taking the curl of the curl, $\partial_t^2$ denotes the second partial time derivative operator, and $\partial_t$ is the partial time derivative operator. The dynamics in (A-6) are separable into the product of a time-independent factor and an exponential $e^{-i\omega t}$ factor where $\omega$ is the angular frequency. This is a consequence of the dynamics being linear and time-invariant. Restricting to physically relevant fields, those which are real-valued and square-integrable, allows any time-dependent solution to be expressed as a sum of monochromatic solutions (those for which $\omega$ is real valued). It is without loss of generality, therefore, that we restrict ourselves to the monochromatic solutions of equation (A-6). Explicitly, these are solutions of the form

$$\boldsymbol{e}(\boldsymbol{r}, t) = \boldsymbol{e}_r(\boldsymbol{r}, \omega) e^{-i\omega t} \tag{A-122}$$

$$\boldsymbol{j}(\boldsymbol{r}, t) = \boldsymbol{j}_r(\boldsymbol{r}, \omega) e^{-i\omega t} \tag{A-123}$$

Where $\boldsymbol{r}$ is the spatial coordinate and the subscript $r$ denotes the time-invariant monochromatic solution. The condition that $\boldsymbol{e}$ and $\boldsymbol{j}$ are real-valued implies

$$\begin{aligned} \boldsymbol{e}_r(\boldsymbol{r}, \omega) &= \boldsymbol{e}_r^*(\boldsymbol{r}, -\omega) \\ \boldsymbol{j}_r(\boldsymbol{r}, \omega) &= \boldsymbol{j}_r^*(\boldsymbol{r}, -\omega) \end{aligned} \tag{A-124}$$

Where the superscript $*$ denotes complex conjugation. This condition allows restriction of $\omega$ to non-negative values only. A monochromatic solution exists for each value of $\omega$. Applying (A-7) and (A-8) onto equation (A-6) and assuming $\omega \neq 0$ (that is, ignoring non-radiating static fields)

$$\boldsymbol{j}_r = \left[ \frac{(\nabla \times)^2 - k_0^2}{i\omega\mu_0} \right] \boldsymbol{e}_r \tag{A-125}$$

where $k_0$ is the wavenumber of the light in vacuum and is defined using

$$k_0 := \frac{\omega}{c} = \omega\sqrt{\epsilon_0\mu_0} \tag{A-126}$$

where $c$ is the speed of light in vacuum. Let us define the linear operator on vector fields

$$\underline{J}_e := \frac{(\nabla \times)^2 - k_0^2}{i\omega\mu_0} \tag{A-127}$$

The underbar _ will be used to denote operators on fields wherever appropriate. For readability, the subscript $r$ will be left implied going forward (except where stated otherwise) so that $\boldsymbol{e}$ will now denote monochromatic solutions. Substituting (A-12) into (A-10)

$$\boldsymbol{j} = \underline{J}_e \boldsymbol{e} \tag{A-128}$$

Equation (A-12) clarifies the role of $\underline{J}_e$. It is an operator that will output the current density field given the electric field. More specifically, the operator returns a microscopic non-displacement electric current density, where displacement current is taken to be the vacuum displacement current

$$\boldsymbol{j}_D := \epsilon_0 \frac{\partial \boldsymbol{e}}{\partial t} \tag{A-129}$$

The vacuum displacement current is removed because it does not account for the movement of charges and therefore should be treated separately. The current density $\boldsymbol{j}$ is particularly significant because it is the only microscopic material response field and thus works at any scale.

Let $\underline{M}$ be an operator which converts from the microscale fields to macroscale fields. $\underline{M}$ encodes the procedure that is used to convert microscopic fields to macroscopic ones. In practice, $\underline{M}$ is often some form of volume averaging operator. We may then express the macroscopic current and field as

$$\begin{aligned} \boldsymbol{J} &= \underline{M}\boldsymbol{j} \\ \boldsymbol{E} &= \underline{M}\boldsymbol{e} \\ \boldsymbol{B} &= \underline{M}\boldsymbol{b} \end{aligned} \tag{A-15}$$

Using equation (A-13)

$$\boldsymbol{J} = \underline{M}\boldsymbol{j} = \underline{M}\underline{J_e}\boldsymbol{e} = \underline{J_e}\underline{M}\boldsymbol{e} = \underline{J_e}\boldsymbol{E} \tag{A-16}$$

Where the micro-to-macro operator $\underline{M}$ and the current density operator $\underline{J_e}$ are assumed to commute. This is true if $\underline{M}$ is linear, space-invariant, and time-invariant (as is often the case – see, for example section 6.6 of [1]). Equation (A-16) reveals that the relationship between current and electric fields described by Maxwell's equations often generalizes to the macroscale.

The constitutive equation relating $\boldsymbol{J}$ to $\boldsymbol{E}$ and $\boldsymbol{B}$ is usually an empirical observation that is based on measurements in the macroscale. Currents induced in an ohmic material are described by the equation

$$\boldsymbol{J} = \sigma\boldsymbol{E} = \underline{J_e}\boldsymbol{E} \tag{A-17}$$

Where the conductivity $\sigma$ is most generally a tensor field relating the electric and current densities at each point. Equation (A-17) sets up an equation for the dynamics in a material. Using our definitions (A-11) and (A-12)

$$0 = [\underline{J_e} - \sigma]\boldsymbol{E} = \left[(\nabla \times)^2 - \left(1 + \frac{i\sigma}{\omega\epsilon_0}\right)k_0^2\right]\boldsymbol{E} \tag{A-18}$$

We can then relate this to the complex relative permittivity $\epsilon$, complex relative permeability $\mu$, and complex refractive index using

$$\mu\epsilon = n^2 = 1 + \frac{i\sigma}{\omega\epsilon_0} \tag{A-19}$$

Which, in equation (A-18), yield the more familiar

$$\begin{aligned} 0 &= [(\nabla \times)^2 - \mu\epsilon k_0^2]\boldsymbol{E} \\ 0 &= [(\nabla \times)^2 - n^2 k_0^2]\boldsymbol{E} \end{aligned} \tag{A-20}$$

[1] J. Jackson. *Classical Electrodynamics.* 3rd Edition, Wiley, New York, NY, (1999)

**Appendix B: Definition for the norm of a linear operator**

The linear operator norm $\|\cdot\|$ is defined for an operator $A: \mathcal{V} \to \mathcal{V}$ acting on a normed vector space $\mathcal{V}$ as smallest real number $a$ which satisfies $|A\boldsymbol{v}| \leq a|\boldsymbol{v}|$ for any vector $\boldsymbol{v}$, where $|\cdot|$ is the vector norm. Formally,

$$\|A\| \coloneqq \inf\{a \in \mathbb{R} | \forall \boldsymbol{v} \in \mathcal{V}: |A\boldsymbol{v}| \leq a|\boldsymbol{v}|\} \tag{B-1}$$

Where inf returns the infimum of the set, and $\mathbb{R}$ is the set of real number. For vector norms induced by the inner product (i.e. $|\boldsymbol{v}| \coloneqq \sqrt{\langle \boldsymbol{v}, \boldsymbol{v} \rangle}$), the operator norm of $\|A\|$ is equivalent to the square root of the largest eigenvalue of $A^\dagger A$ (which is guaranteed to be diagonalizable via the spectral theorem). This is often easier to compute in practice for operators on finite vector spaces, which is the case here. Let $\lambda_{max}(A)$ return the eigenvalue of the operator $A$ with the largest magnitude, then we may write

$$\|A\| = \sqrt{\lambda_{max}(A^\dagger A)} \tag{B-2}$$

If $A$ is diagonalizable, e.g. the permittivity of a passive and lossless media, then

$$\|A\| = \lambda_{max}(A) \tag{B-3}$$

in which case the norm is the eigenvalue with the largest magnitude.

## Appendix C: Conduction in non-absorbing media

If a medium is passive and lossless, then there is no time averaged source or sink of radiation. Mathematically, this corresponds to the time averaged divergence of the Poynting vector being zero,

$$\langle \nabla \cdot \mathbf{S} \rangle_t = 0 \tag{C-1}$$

Taking the time average of equation (C-1) for a monochromatic, real-valued wave and denoting the time-independent complex amplitudes with the subscript $r$ we get:

$$0 = \left\langle \mathbf{E} \cdot \mathbf{J} - i\omega \left( \epsilon_0 E^2 + \frac{B^2}{\mu_0} \right) + \nabla \cdot \mathbf{S} \right\rangle_t = \frac{1}{2} \Re \left( \langle \mathbf{E}_r, \mathbf{J}_r \rangle - i\omega \left( \epsilon_0 E_r^2 + \frac{B_r^2}{\mu_0} \right) \right)$$
$$= \frac{1}{2} \Re \langle \mathbf{E}_r, \mathbf{J}_r \rangle = \frac{1}{4} (\langle \mathbf{E}_r, \mathbf{J}_r \rangle + \langle \mathbf{E}_r, \mathbf{J}_r \rangle^*) = \frac{1}{4} (\langle \mathbf{E}_r, \mathbf{J}_r \rangle + \langle \mathbf{J}_r, \mathbf{E}_r \rangle) \tag{C-2}$$

Where $\Re$ returns the real component of a complex number, the superscript $*$ denotes complex conjugation, and $\langle \cdot, \cdot \rangle$ denotes the inner product defined in any orthonormal basis as

$$\langle \mathbf{A}, \mathbf{B} \rangle \coloneqq \sum_i A_i^* B_i \coloneqq \mathbf{A}^\dagger \mathbf{B} \tag{130}$$

Where $A_i$ is the $i$th component of $\mathbf{A}$ and $B_i$ the $i$th component of $\mathbf{B}$. In general, equation (C-2) reveals that the monochromatic solution for a passive and lossless media satisfies

$$\langle \mathbf{E}, \mathbf{J} \rangle = -\langle \mathbf{J}, \mathbf{E} \rangle \tag{C-131}$$

Where we have left the subscript $r$ implied again. Note that since any time-dependent solution is a superposition of time independent solutions, equation (C-4) is still true for a time-dependent solution. If the constitutive equation (A-17) is also satisfied (material is linear and current is dominated by electric driving force), then

$$\mathbf{E}^\dagger \sigma \mathbf{E} = \mathbf{E}^\dagger \mathbf{J} = \langle \mathbf{E}, \mathbf{J} \rangle = -\langle \mathbf{J}, \mathbf{E} \rangle = -(\sigma \mathbf{E})^\dagger \mathbf{E} = -\mathbf{E}^\dagger \sigma^\dagger \mathbf{E} \tag{132}$$

Where $\sigma$ is in general a tensor and the superscript † denotes the adjoint, or the conjugate transpose of the representative matrix in an orthogonal basis. Since equation (C-5) is true for any choice of $\mathbf{E}$, the conductivity tensor $\sigma$ for a passive, lossless medium must satisfy

$$\sigma^\dagger = -\sigma \tag{133}$$

Furthermore, the proof can be run backwards to show that a conductivity tensor satisfying (C-6) is sufficient to show that (C-1) is satisfied. That is, condition (C-6) is equivalent to being a linear, passive, and lossless material. Using definition (A-19), condition (C-6) can be restated as a condition on the other material properties

$$n^2 = (n^2)^\dagger \tag{134}$$
$$\mu\epsilon = (\mu\epsilon)^\dagger \tag{135}$$

If the material is non-magnetic (or more generally, has a weakly anisotropic, Hermitian magnetic response), as is often the case in dielectrics then condition (C-8) becomes

$$\epsilon = \epsilon^\dagger \tag{136}$$

In all four conditions (C-6)-(C-9), the tensor in question is normal. Applying the spectral theorem from linear algebra, this implies that:

1. The tensor is diagonalizable
2. Its eigenvectors are orthogonal to each other
3. The tensor can be diagonalized by a rotation

Often, it is mathematically expedient to represent tensors using a diagonal matrix. Conditions (46)-(49) guarantees that a linear, passive, and lossless medium has electromagnetic material properties that are diagonalizable using a rotation. In practice, whenever possible the response tensor is diagonalized and the eigenvectors are called the principal axis. In the case of the $n^2$ response tensor, the eigenvalues are the squares of the principal refractive indices.

**Appendix D: Intensity of unidirectional light in media**

Since our calculations depend on intensity being proportional to $|E|^2$ in weakly anisotropic far field media, we wish to show that such media will exhibit intensity proportional to $|E|^2$. The microscopic intensity (irradiance) through an infinitesimal surface at $r$ is the time-averaged flux through that surface and can be written as

$$i(r) = |\langle s(r,t) \rangle_t| \cos \beta \tag{D-1}$$

Where $\langle \cdot \rangle_t$ is a time average and $\beta$ is the angle between $S$ and the surface normal. The time-dependent Poynting vector is

$$s(r,t) := \frac{1}{\mu_0} e(r,t) \times b(r,t) \tag{D-2}$$

Where $e(r,t)$ is the time-dependent electric field, and $b(r,t)$ is the time-dependent magnetic field. Assuming that the medium is time invariant, then the monochromatic solution (7) may be used. Equation (A-7) substituted in Equation (A-3) implies that the magnetic field $b(r,t)$ is also monochromatic. Time averaging of monochromatic fields in equation (D-2) and substituting into equation (D-1) yields

$$i(r, \omega) = \frac{1}{2\mu_0} |\Re[e(r,\omega) \times b^*(r,\omega)]| \cos \beta \tag{D-3}$$

where $e(r,\omega)$ and $b(r,\omega)$ now represent the time-independent phasors at $r$ for frequency $\omega$. The intensity of a solution that is the sum of multiple monochromatic modes can be found by summing the intensities of each mode. This is because the cross terms (e.g. $e(r,\omega_1) \times b^*(r,\omega_2)$ where $\omega_1 \neq \omega_2$) always vanish when time averaged over a long enough interval. To illustrate

$$\langle e(r,\omega_1) \times b^*(r,\omega_2) \rangle_t = e_r(r,\omega_1) \times b_r(r,\omega_2) \left( \lim_{T \to \infty} \frac{1}{T} \int_{-T/2}^{T/2} e^{i(\omega_2 - \omega_1)t} dt \right)$$
$$= \begin{cases} 0 & \omega_1 \neq \omega_2 \\ 1 & \omega_1 = \omega_2 \end{cases} \tag{D-4}$$

If the time average is taken over a finite interval cross terms between similar frequencies may not vanish. In practice instruments used to measure intensity do not take an infinitely long average, so that $T$ is finite and the condition

$$\omega_2 - \omega_1 \gg \frac{2\pi}{T} \tag{D-5}$$

is sufficient to ensure that the cross terms do not appear in the measurement. This places a fundamental restriction on the frequency resolution an instrument that has a finite dwell time $T$. Here we will focus on an ideal intensity where $T \to \infty$ and the cross terms do not appear in the calculation. This allows us to focus only on the intensity of monochromatic waves without regard to dispersion effects on intensity.

To find the intensity using equation (D-3), we need to know the relationship between $\boldsymbol{e}$ and $\boldsymbol{b}$. Applying monochromatic electric fields to a linear, time-invariant medium in (A-3) yields

$$\boldsymbol{b} = \frac{\nabla \times \boldsymbol{e}}{i\omega} \tag{D-6}$$

Equation (D-6) can be used in the time averaged Poynting vector to yield

$$\begin{aligned}
\langle \boldsymbol{s} \rangle_t &= \frac{|\Re[\boldsymbol{e} \times \boldsymbol{b}^*]|}{2\mu_0} = \frac{|\Im[\boldsymbol{e}^* \times (\nabla \times \boldsymbol{e})]|}{2\mu_0 \omega} \\
&= \frac{\left|\Im\left[e^2 \left(\hat{\boldsymbol{e}}^* \times \left(\frac{\nabla e}{e} \times \hat{\boldsymbol{e}}\right)\right) + e^2(\hat{\boldsymbol{e}}^* \times (\nabla \times \hat{\boldsymbol{e}}))\right]\right|}{2\mu_0 \omega} \\
&= \frac{e^2}{2\mu_0 \omega} |\Im[\hat{\boldsymbol{e}}^* \times (\nabla \ln e \times \hat{\boldsymbol{e}}) + \hat{\boldsymbol{e}}^* \times (\nabla \times \hat{\boldsymbol{e}})]|
\end{aligned} \tag{D-7}$$

Substituting into (D-3) we get

$$\begin{aligned}
i(\boldsymbol{r}, \omega) &= \frac{\cos \beta}{2\mu_0 \omega} |\Im[\boldsymbol{e}^* \times (\nabla \times \boldsymbol{e})]| \\
&= \frac{E^2 \cos \beta}{2\mu_0 \omega} |\Im[\hat{\boldsymbol{e}}^* \times (\nabla \ln e \times \hat{\boldsymbol{e}}) + \hat{\boldsymbol{e}}^* \times (\nabla \times \hat{\boldsymbol{e}})]|
\end{aligned} \tag{D-8}$$

Equation (D-8) provides a way to calculate the microscopic light intensity from the electric fields of any monochromatic mode.

An Ohmic medium which is uniform on the scale of a wavelength $\lambda_m$ will be able to support unidirectional waves with wavenumber $k \coloneqq 2\pi/\lambda_m$. Since we require a uniformity on the scale of the wavelength, microscopic detail on a scale smaller than the wavelength can be discarded. Thus, unidirectionality can be regarded as a macroscopic phenomenon and much of what follows will only apply when the material's macroscopic properties allow unidirectional wave solutions. Usually, there are many micro-to-macro operators $\underline{M}$ that will work. The mathematical requirements for a valid micro-to-macro operator are that it needs to commute with the current-density operator (see **appendix A**), it allows modes relevant to scattering (e.g. for elastic scattering the Ewald sphere) to be recovered, it ensures the macroscopic fields satisfy a uniform ohmic constitutive equation, and it allows recovery of intensity modes having wavenumbers of interest (e.g. for elastic scattering, intensity modes with wavenumber $2k$). For a linearly polarized, unidirectional electromagnetic wave propagating with wavevector $\boldsymbol{k} = k\hat{\boldsymbol{k}}$, equation (D-8) becomes

$$I(\boldsymbol{r}, \omega) = \frac{k(\boldsymbol{r}, \omega)E^2 \cos \beta}{2\mu_0 \omega} |\hat{\boldsymbol{E}} \times \hat{\boldsymbol{k}} \times \hat{\boldsymbol{E}}| = \frac{\cos \beta}{2\mu_0 c} n(\boldsymbol{r}, \hat{\boldsymbol{k}}, \omega) \sin \alpha(\boldsymbol{r}, \hat{\boldsymbol{k}}, \omega) E^2 \tag{D-9}$$

Where $n$ is the index of refraction (a macroscopic property, note that only $n^2$ was defined to be microscopic) and $\alpha$ is the angle between $\boldsymbol{k}$ and $\boldsymbol{E}$. Equation (D-9) is a general equation for a monochromatic, unidirectional intensity. Further simplifications can be made for a weakly anisotropic medium. A medium is weakly anisotropic if it satisfies

$$\|n^2 - \langle n^2 \rangle_\Omega\| \ll |\langle n^2 \rangle_\Omega| \tag{D-10}$$

Where $\langle \cdot \rangle_\Omega$ is an average taken along all orientations, and $\|\cdot\|$ is the linear operator norm as described in **appendix B**. A simple way to compute the orientation average of a tensor is to use

$$\langle n^2 \rangle_\Omega = \frac{1}{3} \text{tr}(\mu\epsilon) \tag{D-11}$$

where $\text{tr}(\cdot)$ is the trace of an operator. In a lossless, Ohmic medium, the weak anisotropy condition allows us to use the approximations

$$\begin{aligned} n^2 &\approx \bar{n}^2 \\ n &\approx \bar{n} \end{aligned} \tag{D-12}$$

So that the anisotropic refractive index $n$ can be approximated by an averaged isotropic refractive index $\bar{n}$. It can also be shown that condition (D-10) implies $\sin\alpha\left(\hat{\boldsymbol{k}}, \omega\right) \approx 1$. To show this, begin with the monochromatic, unidirectional, and Ohmic version of (4)

$$i\boldsymbol{k} \times \boldsymbol{B} = (\mu_0\sigma - i\omega\mu_0\epsilon_0)\boldsymbol{E} = -i\omega\mu_0\epsilon_0 n^2 \boldsymbol{E} \tag{D-13}$$

Taking the dot product of (D-13) with $\hat{\boldsymbol{k}}$

$$0 = \hat{\boldsymbol{k}} \cdot (n^2 \boldsymbol{\hat{E}}) = \bar{n}^2 \hat{\boldsymbol{k}} \cdot \boldsymbol{\hat{E}} + \hat{\boldsymbol{k}} \cdot \left((n^2 - \bar{n}^2)\boldsymbol{\hat{E}}\right) \tag{D-14}$$

Rearranging and using Cauchy-Swartz inequality

$$\left|\hat{\boldsymbol{k}} \cdot \boldsymbol{\hat{E}}\right| = \left|\frac{\hat{\boldsymbol{k}} \cdot \left((n^2 - \bar{n}^2)\boldsymbol{\hat{E}}\right)}{\bar{n}^2}\right| \leq \frac{|(n^2 - \bar{n}^2)\boldsymbol{\hat{E}}|}{|\bar{n}^2|} \tag{D-15}$$

Since the far field medium is lossless, $n^2$ is Hermitian and has orthogonal eigenvectors with eigenvalues corresponding to $n_1^2, n_2^2, n_3^2$ in the principal axes. The tensor $(n^2 - \bar{n}^2)$ is Hermitian with eigenvalues corresponding to $n_1^2 - \bar{n}^2, n_2^2 - \bar{n}^2, n_3^2 - \bar{n}^2$. Since all its eigenvalues are finite, the tensor $(n^2 - \bar{n}^2)$ is bound and, using (54), its norm is

$$\|n^2 - \bar{n}^2\| = \max(|n_i^2 - \bar{n}^2|) \tag{D-16}$$

So that equation (D-15) becomes

$$\left|\hat{\boldsymbol{k}} \cdot \boldsymbol{\hat{E}}\right| \leq \frac{\|n^2 - \bar{n}^2\||\boldsymbol{\hat{E}}|}{|\bar{n}^2|} \leq \frac{\|n^2 - \bar{n}^2\|}{|\bar{n}^2|} \ll 1 \tag{D-17}$$

Where the weak anisotropy condition (D-10) in the far-field was used in the final inequality. This implies that

$$\sin\alpha = \sqrt{1 - \left|\hat{\boldsymbol{k}} \cdot \boldsymbol{\hat{E}}\right|^2} \approx 1 \tag{D-18}$$

Using (D-13) and (D-18) Equation (D-9) can, whenever the far-field medium satisfies (D-10), be approximated by

$$I(\boldsymbol{r}, \omega) = \frac{\bar{n}\cos\beta}{2\mu_0 c} |\boldsymbol{E}(\boldsymbol{r}, \omega)|^2 \coloneqq G|\boldsymbol{E}(\boldsymbol{r}, \omega)|^2 \tag{D-19}$$

Where $G$ is a proportionality constant that depends on the medium and the angle $\beta$ between the surface normal and the Poynting vector. Equation (D-19) is the intensity for a monochromatic, unidirectional mode in a weakly anisotropic medium. An intensity measurement usually measures the total power passing through a surface $A$

$$P(\omega) = \iint_A d\mathbf{r}\, I(\mathbf{r}, \omega) = \iint_A d\mathbf{r}\, G(\mathbf{r}) |\mathbf{E}(\mathbf{r}, \omega)|^2 \qquad \text{(D-20)}$$

If the collecting surface is designed such that $\beta(\mathbf{r})$ is constant (e.g. by making $\beta = 0$ everywhere or using a flat surface small enough that the direction of $\mathbf{S}$ is uniform throughout), then the measured power becomes

$$P(\omega) = G \iint_A d\mathbf{r}\, |\mathbf{E}(\mathbf{r}, \omega)|^2 \qquad \text{(D-21)}$$

In such cases the actual value of $G$ is unimportant since 1) a measured value of $I$ is usually in units of voltage or counts so that an unknown proportionality constant persists when relating theory to data and 2) it is divided out when calculating transmission. Incidentally, equation (D-20) is the same as would be derived for an isotropic medium whose refractive index is $\bar{n}$. Recalling our assumptions, equation (D-21) is satisfied whenever the radiation is unidirectional, the temporal averaging is sufficiently large to eliminate cross terms, the medium is ohmic, macroscopically uniform, and weakly anisotropic.